\documentclass[prd,aps,twocolumn,showkeys,nofootinbib]{revtex4-2}

\usepackage{amsmath}
\usepackage{amsfonts}
\usepackage{amssymb}	
\usepackage{graphicx}
\usepackage{bm}
\usepackage{color}
\usepackage{ulem}
\usepackage{commath}
\usepackage[T1]{fontenc}
\usepackage{lmodern}
\usepackage{xcolor}
\usepackage{hyperref}
\allowdisplaybreaks

\def\GMc2{G M_{\odot} c^{-2}}

\def\lm{{\ell m}}

\def\lm{{\ell m}}

\def\de{\partial}
\def\lm{{\ell m}}

\def\ph{\varphi}

\newcommand\be{\begin{equation}}
\newcommand\ee{\end{equation}}

\def\Qo{Q_{\omega}}
\def\FISF{\mathcal{F}_\ph^{\rm 1SF}}
\def\FIISF{\mathcal{F}_\ph^{\rm 2SF}}
\def\FIIISF{\mathcal{F}_\ph^{\rm 3SF}}

\def\TEOBResumS{\texttt{TEOBResumS}}

\def\TEOBResumSDali{\texttt{TEOBResumS-DALI}}

\def\PAT{1PAT1}

\begin{document}

\title{Comparing second-order gravitational self-force and effective one body
waveforms from inspiralling, quasi-circular and nonspinning black hole binaries II: 
the large-mass-ratio case}
\author{Angelica \surname{Albertini}${}^{1,2}$}
\author{Alessandro \surname{Nagar}${}^{3,4}$}
\author{Adam \surname{Pound}${}^5$}
\author{Niels \surname{Warburton}${}^6$}
\author{Barry \surname{Wardell}${}^6$}
\author{Leanne \surname{Durkan}${}^6$}
\author{Jeremy \surname{Miller}${}^7$}

\affiliation{${}^1$Astronomical Institute of the Czech Academy of Sciences,
Bo\v{c}n\'{i} II 1401/1a, CZ-141 00 Prague, Czech Republic}
\affiliation{${}^2$Faculty of Mathematics and Physics, Charles University in Prague, 18000 Prague, Czech Republic}
\affiliation{${}^3$INFN Sezione di Torino, Via P. Giuria 1, 10125 Torino, Italy} 
\affiliation{${}^4$Institut des Hautes Etudes Scientifiques, 91440 Bures-sur-Yvette, France}
\affiliation{${}^5$School of Mathematical Sciences and STAG Research Centre, University of Southampton, Southampton, United Kingdom, SO17 1BJ}
\affiliation{${}^6$School of Mathematics and Statistics, University College Dublin, Belfield, Dublin 4, Ireland, D04 V1W8}
\affiliation{${}^7$Department of Physics, Ariel University, Ariel 40700, Israel}


\begin{abstract}
We compare recently computed waveforms from second-order gravitational self-force (GSF) theory to 
those generated by a new, GSF-informed, effective one body (EOB) waveform model for (spin-aligned, eccentric) 
inspiralling black hole binaries with large mass ratios. We focus on quasi-circular, nonspinning, 
configurations and perform detailed  GSF/EOB waveform phasing comparisons, either in the time domain 
or via the gauge-invariant dimensionless function $Q_\omega\equiv \omega^2/\dot{\omega}$, 
where $\omega$ is the gravitational wave frequency. The inclusion of high-PN test-mass terms within 
the EOB  radiation reaction (notably, up to 22PN) is crucial to achieve an EOB/GSF phasing agreement  
below 1~rad up to the end of the inspiral for mass ratios up to 500. For larger mass ratios, 
up to $5\times 10^4$,  the contribution of horizon absorption becomes more and more important 
and needs to be accurately modeled. Our results indicate that our GSF-informed EOB waveform model
is a promising tool to describe waveforms generated by either intermediate or extreme mass 
ratio inspirals for future gravitational wave detectors.
\end{abstract}

\date{\today}

\maketitle

\section{Introduction}

In the next twenty years we will witness the development and the launch of new gravitational wave
observatories, both ground-based, like Einstein Telescope (ET)~\cite{Punturo:2010zz, Maggiore:2019uih}
and Cosmic Explorer (CE)~\cite{Reitze:2019iox, Evans:2021gyd}, and space-based, like LISA~\cite{LISA:2017pwj},
TianQuin~\cite{TianQin:2015yph} and Taiji~\cite{Ruan:2018tsw}. 
Given their increased sensitivities and the larger frequency range they will cover, these detectors will be able 
to see many more sources than the compact binary coalescences currently observed by the LIGO-Virgo-KAGRA collaboration.
In particular, they are expected to observe both intermediate- and extreme-mass-ratio inspirals 
(IMRIs and EMRIs, respectively) with mass ratios ranging between $10^{-2}-10^{-4}$ in the first
case, and $\sim 10^{-4}-10^{-7}$ in the second case. EMRIs in particular are binaries where a stellar-mass
compact object inspirals for years around a supermassive black hole. The waveform phenomenology
for these binaries can be extremely complicated, as it may involve at the same time both high eccentricity 
and rapid precession of the orbital plane. Because of the resulting, rich harmonic structure in the waveform, and the large number of orbits near merger, these extreme-mass-ratio inspirals have the potential to 
unveil and test deep features of strong-field General Relativity~\cite{Berry:2019wgg}.
Although the dynamics of the binary can be seen as a perturbation to the underlying Kerr spacetime, 
the accurate description of its evolution is challenging, since the self-field of the smaller object cannot be neglected.
The accurate description of the action of the self-field of the smaller object on the dynamics
is described within the gravitational self-force (GSF) program~\cite{Pound:2015tma, Pound:2017psq, Barack:2018yvs}. 
There are different efforts in developing waveform templates for EMRIs building on GSF
results~\cite{Sundararajan:2008zm, Pound:2015tma,Barack:2018yvs,Miller:2020bft,Hughes:2021exa,Pound:2021qin,
Wardell:2021fyy,Warburton:2021kwk, Hughes:2021exa}. Much less accurate, though faster schemes, 
go under the name of kludge waveforms~\cite{Barack:2003fp,Babak:2006uv, Chua:2015mua, Chua:2017ujo}. 
The recently released \texttt{FastEMRIWaveform} package~\cite{Chua:2020stf, Katz:2021yft}
combines speed and accuracy, but is currently implemented only for eccentric inspirals into a non-rotating black hole.
All these approaches are based on expansions in the mass ratio, under the hypothesis that it is small.

The effective one body (EOB) approach~\cite{Buonanno:1998gg,Buonanno:2000ef,Damour:2000we,Damour:2001tu,Damour:2015isa}, on the other hand,  
is a powerful analytical framework that can describe the dynamics of 
any binary all over the parameter space, for any value of the mass ratio and for any orientation of the
vectorial individual spins. Although the EOB method has been largely exploited in building templates 
for comparable-mass binaries,  it is a natural framework to construct waveforms also for binaries with 
large mass ratios~\cite{Yunes:2009ef,Yunes:2010zj, Albanesi:2021rby}, since it builds upon a deformation 
of the geometry of a Kerr black hole with the symmetric mass ratio as the deformation parameter.

In a recent work~\cite{Albertini:2022rfe} (hereafter Paper~I), we compared waveforms generated 
by the state-of-the-art EOB model \TEOBResumS{}~\cite{Nagar:2018zoe, Nagar:2019wds, Nagar:2020pcj,Riemenschneider:2021ppj},
with those from a complete gravitational self-force model called \PAT{}~\cite{Wardell:2021fyy}.
Although the \PAT{} model is crucially lacking the transition from inspiral to plunge,
it was possible to show  that the major difference between the two models arises from
contributions that  are {\it linear} in the symmetric mass ratio $\nu\equiv m_1 m_2/M^2$,
where $m_{1,2}$ are the masses of the two bodies, $M\equiv m_1+m_2$ and we use 
the convention $m_1\geq m_2$. 
This result is not surprising. Since \TEOBResumS{} was originally conceived as a waveform model 
aiming at primarily generating waveforms for comparable mass binaries\footnote{Note however that the model 
was found to perform consistently with NR also for large mass ratios, up to the intermediate 
mass-ratio regime~\cite{Nagar:2022icd}.}, a relatively limited amount of linear-in-$\nu$ (or test-mass) 
information (both in the conservative and nonconservative sector of the model) was included. 
The rationale behind this choice was to use some test-mass information to improve the behavior of the
model for comparable masses, while avoiding this information becoming dominant in this case.
A typical example of this procedure is that \TEOBResumS{} partly relies  on a mixed $3^{+2}$PN 
and $3^{+3}$PN resummed flux~\cite{Damour:2009kr}; i.e., full $\nu$-dependent terms up to 3PN 
are {\it hybridized} with test-mass terms up to 5PN or 6PN depending on the multipole. These hybrid 
expressions are further resummed using various choices of Pad\'e approximants~\cite{Nagar:2019wds, Nagar:2020pcj}.
Similarly, the \TEOBResumS{} Hamiltonian only incorporates terms up to 3PN or 5PN, 
depending on the particular EOB potential.

A step towards incorporating full 1GSF information (i.e., linear in $\nu$) in the Hamiltonian 
was taken by Antonelli et al.~\cite{Antonelli:2019fmq}. In particular, they used a post-Schwarzschild 
Hamiltonian~\cite{Damour:2017zjx,Bini:2020wpo} to overcome the well-known 
problems related to the the presence of the light-ring coordinate singularity in the
standard EOB gauge (or Damour-Jaranowski-Sch\"afer, DJS hereafter~\cite{Damour:2000we})~\cite{Akcay:2012ea}. 
Although promising, the approach  of~\cite{Antonelli:2019fmq}, that was limited to the case of 
nonspinning binaries, needs more development to construct a complete model, 
informed by Numerical Relativity simulations, able to span the full range of mass ratios.
Reference~\cite{Nagar:2022fep} introduced an alternative strategy, that was however 
specifically designed for IMRIs and EMRIs (including aligned-spin and eccentricity),
where the importance of the merger is practically negligible and the signal-to-noise ratio
is dominated by the hundred of thousands of cycles of the inspiral. To target these 
sources, Ref.~\cite{Nagar:2022fep} proposed an EOB model in the DJS gauge 
(and thus with the well-known light-ring singularity) but introduced suitable resummation 
procedures to improve the behavior of the PN-expanded EOB potentials in the strong field. 
By additionally informing them with exact GSF results~\cite{Akcay:2015pjz}, 
Ref.~\cite{Nagar:2022fep} introduced the first, and so far only, EOB waveform model for 
eccentric, spin-aligned IMRIs and EMRIs that is informed by GSF numerical results.

The aim of this paper is to test the GSF-informed EOB model of~\cite{Nagar:2022fep} 
against the 2GSF waveforms of~\cite{Wardell:2021fyy}, analogously to what 
Paper~I did using the standard \TEOBResumS{} model. 
The paper is organized as follows. In Sec.~\ref{sec:GSF-EOB} we recap the main elements
of the model of~\cite{Nagar:2022fep}, reminding readers of the structure of the Hamiltonian and giving
some details about the structure of the radiation reaction, which is a novelty  
introduced here. In Sec.~\ref{sec:phasing} we compare the EOB and GSF models, first 
performing a waveform alignment in the time domain and then using the same gauge-invariant 
frequency-domain analysis we exploited in Paper~I.
Section~\ref{Qomg:anlyt} gives a more precise analytic interpretation of the results 
presented in Sec.~\ref{sec:phasing}, while Sec.~\ref{sec:horizon} focuses on the impact
of horizon absorption and on the need to accurately model it within EOB 
to correctly describe EMRIs. Our conclusions are collected in Sec.~\ref{sec:conclusions}
We use units with $G=c=1$ and define the mass ratio $q \equiv m_1/m_2 \ge 1$.

\section{GSF-informed EOB model}
\label{sec:GSF-EOB}

\subsection{The Hamiltonian: a reminder}
Let us briefly recall the elements of the GSF-informed EOB model of Ref.~\cite{Nagar:2022fep}.
The model builds upon the spin-aligned, eccentric \TEOBResumS{} model, 
the \TEOBResumSDali{}~\cite{Chiaramello:2020ehz,Nagar:2021gss,Bonino:2022hkj}, but the low-PN accurate, 
NR-informed EOB potentials are replaced by the 1GSF-informed ones. 
More precisely, the potentials $(A,\bar{D},Q)$ at linear order in $\nu$ formally read
\begin{align}
A &= 1 - 2 u + \nu a_{\rm 1SF}(u)\ , \\
\bar{D} &= 1 + \nu \bar{d}_{\rm 1SF}(u)\ , \\
Q         &=\nu q_{\rm 1SF}(u) \ .
\end{align}
where $u\equiv 1/r = M/R$ is the dimensionless Newtonian potential. Reference~\cite{Nagar:2022fep}, 
building upon 8.5PN results~\cite{Bini:2014nfa}, showed that the 
strong-field (i.e., nearby the last stable orbit) behavior of the three functions $(a_{\rm 1SF},\bar{d}_{\rm 1SF},q_{\rm 1SF})$ 
can be improved by implementing a certain factorization and resummation procedure based on
Pad\'e approximants. Moreover, these Pad\'e-resummed functions can be modified by a 
certain {\it flexing} factor, which effectively takes into account higher-order corrections and
can be informed by fitting to the numerical data of Refs.~\cite{Akcay:2012ea,Akcay:2015pjz}. 
This correcting factor yields GSF-informed analytic potentials that display $\lesssim 0.1\%$ fractional
difference with the numerical data up to $u=1/3$ for $a_{\rm 1SF}$ and up to $u=1/5$ for $\bar{d}_{\rm 1SF}$
and $q_{\rm 1SF}$; see Figs.~2,~3 and ~4 of Ref.~\cite{Nagar:2022fep}. The potentials then enter the
Hamiltonian as described in Sec.~II of~\cite{Nagar:2022fep}.

\subsection{Waveform and radiation reaction}
\label{sec:fluxes}
To fix conventions, the strain waveform is decomposed on spin-weighted spherical harmonics as
\be
h_+ - i h_\times = \dfrac{1}{D_L}\sum_{\ell=2}^{\ell_{\rm max}} \sum_{m=-\ell}^{\ell}h_\lm{}_{-2}Y_\lm(\iota,\phi) \ ,
\ee
where $D_L$ indicates the luminosity distance, ${}_{-2}Y_\lm(\iota,\phi)$ are the $s=-2$
spin-weighted spherical harmonics, $\iota$ is the inclination angle with respect to the orbital plane, 
and $\phi$ the azimuthal one. In the following, we will also work with the Regge-Wheeler-Zerilli 
(RWZ) normalization convention and express the waveform as $\Psi_\lm\equiv h_\lm/\sqrt{(\ell+2)(\ell+1)\ell(\ell-1)}$.
The RWZ normalized strain quadrupole waveform is then separated into amplitude and phase with the convention
\be
\label{eq:RWZnorm}
\Psi_{22} (t)= A_{22}(t) e^{-i \phi_{22}(t)},
\ee
where $t\equiv T/M$ is the time in units of the total mass $M$. The instantaneous
gravitational wave frequency (in units of $M$) is defined as $\omega_{22} \equiv \dot{\phi}_{22}$.
Following~\cite{Damour:2008gu} the waveform multipoles are factorized as $h_\lm = h^{\rm Newt}_\lm \hat{h}_\lm$, 
where the first contribution is the leading, Newtonian one, and the PN corrections is written as
\be
\label{eq:hlm}
\hat{h}_\lm = \hat{S}^{(\epsilon)}_{\rm eff} \, T_\lm e^{i \delta_\lm}  \left( \rho_\lm \right)^{\ell} ,
\ee
where $\epsilon=(0,1)$ is the parity of $\ell + m$, $\hat{S}^{(\epsilon)}_{\rm eff}$ is the effective 
source of the field (effective energy or Newton-normalized angular momentum depending on the parity
of the mode~\cite{Damour:2008gu}), $T_\lm$ is the tail factor, which resums an infinite number of
leading-order logarithms, while $\rho_\lm$ and $\delta_\lm$ are the residual amplitude and phase
corrections, respectively.

For simplicity, Ref.~\cite{Nagar:2022fep} used the standard radiation reaction implemented in
\TEOBResumS{}, with the (resummed) PN orders of the various multipoles chosen as in 
Refs.~\cite{Nagar:2019wds,Nagar:2020pcj}. However, Ref.~\cite{Nagar:2022fep} already
pointed out that the standard \TEOBResumS{} analytical flux needs to be improved 
to achieve a  faithful representation of the exact flux (obtained numerically) in the test-mass limit 
(see in particular Fig.~8 of~\cite{Nagar:2022fep}). Before discussing this issue in some detail,
let us also remember that \TEOBResumS{} also implements horizon absorption following the
prescription of Ref.~\cite{Damour:2014sva}. The horizon flux we are using here only has the 
$\ell =m =2$ and $\ell=2$, $m=1$ modes, following Ref.~\cite{Nagar:2011aa}. In the 
nonspinning case, we have both $\rho_{22}^H$ and $\rho_{21}^H$, implemented as described 
in Ref.~\cite{Bernuzzi:2012ku}.

As a first attempt, we took precisely the model of Ref.~\cite{Nagar:2022fep}, in the quasi-circular
limit\footnote{In particular, setting the radial radiation reaction force to zero, ${\cal F}_r=0$, as is
the case for the native quasi-circular \TEOBResumS{} model.} and performed phasing comparisons
(either in the time domain or using the $\Qo$ function) for different mass ratios up to $q=5000$.
The dephasing we found is largely nonnegligible as $q$ increases, as illustrated in Appendix~\ref{sec:flux}. 
This is not surprising, and it is consistent with the relatively poor accuracy of the standard \TEOBResumS{}
flux in the test-mass limit, as pointed out in the Appendix of~\cite{Nagar:2022fep}.
To overcome this difficulty, we attempt here to use the $3^{+19}$PN radiation reaction
already exploited in Sec.~VA of Ref.~\cite{Nagar:2022icd}. We remind the reader that the notation $3^{+19}$PN
means that the standard 3PN-accurate terms in the $\rho_\lm$'s (that depend on $\nu$) are {\it hybridized}
with test-mass terms (that are $\nu$-independent) so as to achieve global 22PN accuracy~\cite{Fujita:2012cm} 
for all $\rho_\lm$ functions. For simplicity, we do not attempt any additional resummation (e.g., using Pad\'e
approximants) on these resulting hybrid functions, although it might be useful to further improve the
behavior of the residual PN series in strong field, especially in the presence of spin~\cite{Nagar:2016ayt}.
From now on, the $3^{+19}$PN-accurate radiation reaction will become our standard choice
and we will generically refer to it as the {\it hybrid flux}. We will see in the next section that it is essential
to deliver an excellent EOB/GSF phasing agreement for large-mass-ratio binaries.

\begin{figure}[t]
\center
\includegraphics[width=0.43\textwidth]{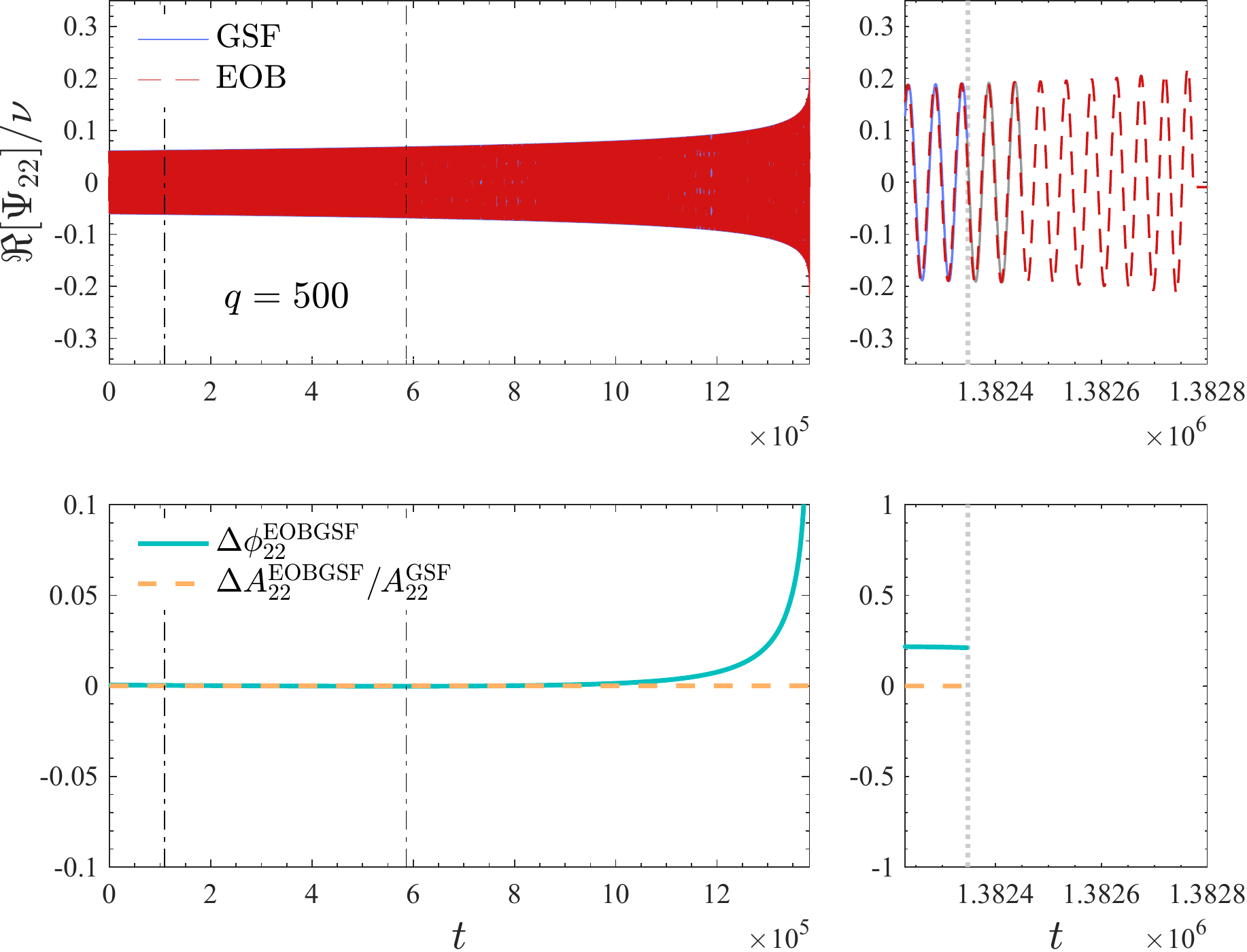} 
\caption{\label{fig:phasing500}EOB/GSF phasing for a $q = 500$ binary. 
The vertical dash-dotted lines in the left panel indicate the times corresponding to the $[\omega_L,\omega_R]$ 
alignment interval. In the right panel the dotted line corresponds to $\omega_{22}^{\rm GSF_{\rm break}}$.}
\end{figure}

\begin{figure*}[t]
\center
\includegraphics[width=0.32\textwidth]{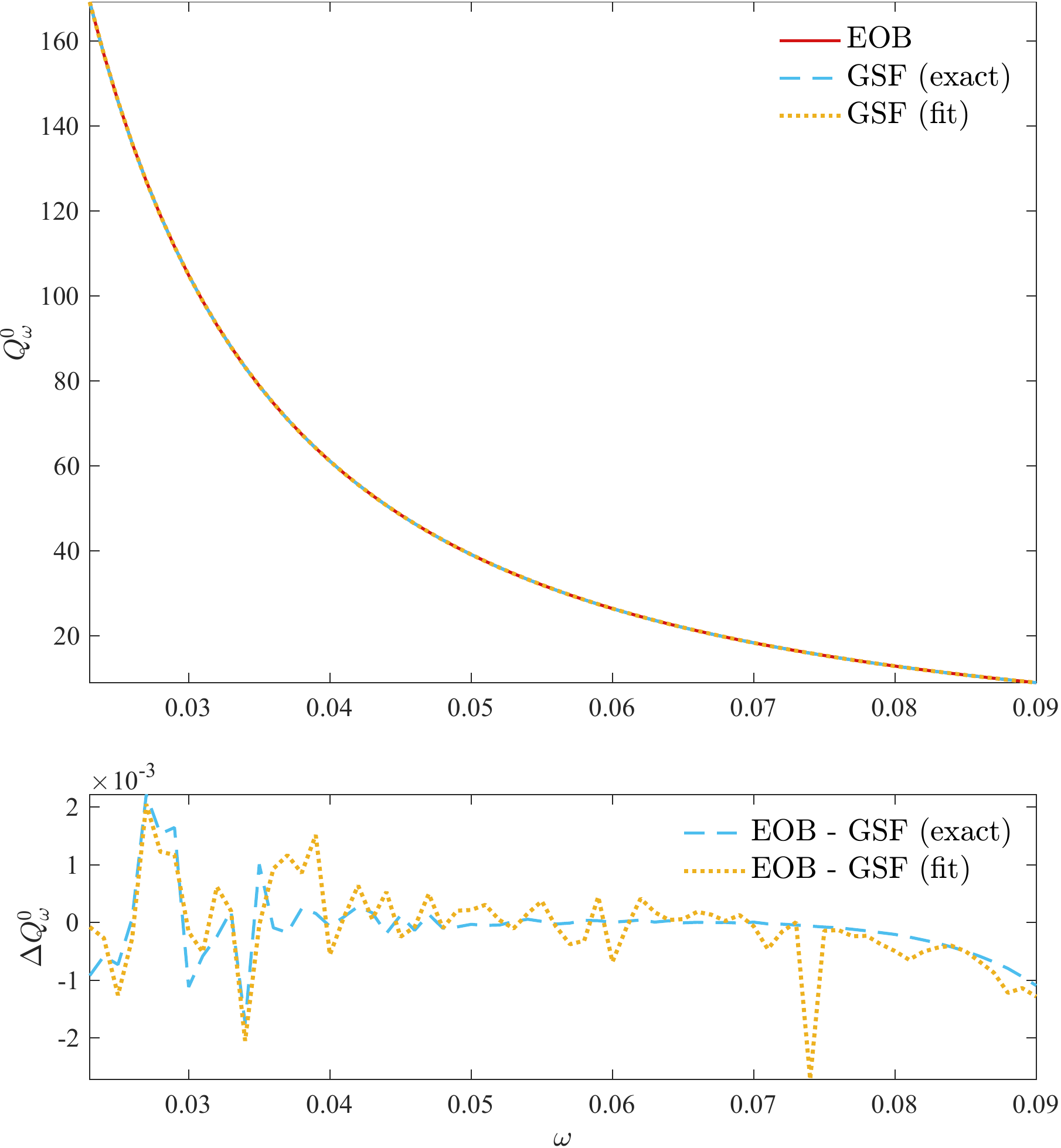} 
\includegraphics[width=0.32\textwidth]{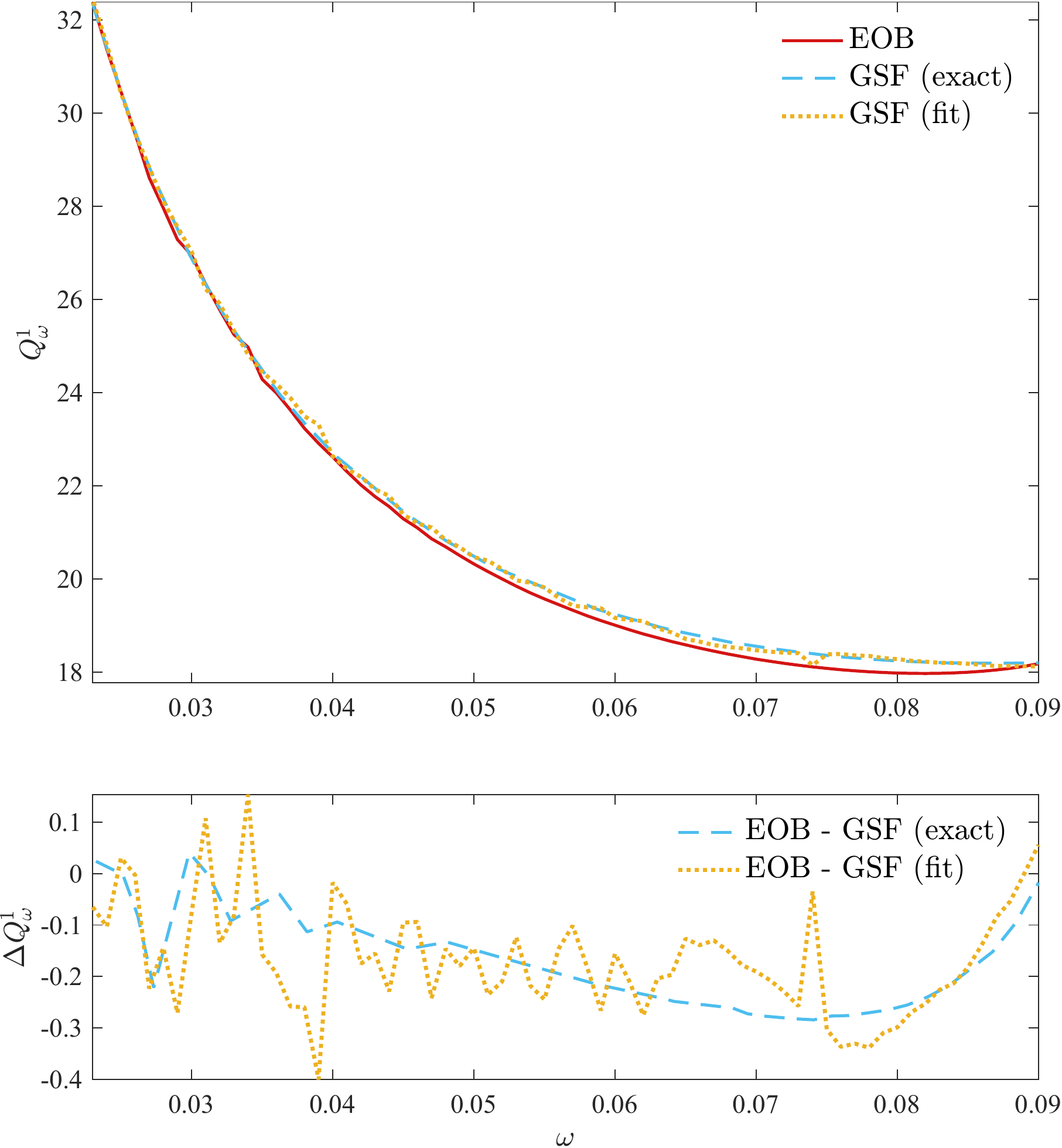} 
\includegraphics[width=0.32\textwidth]{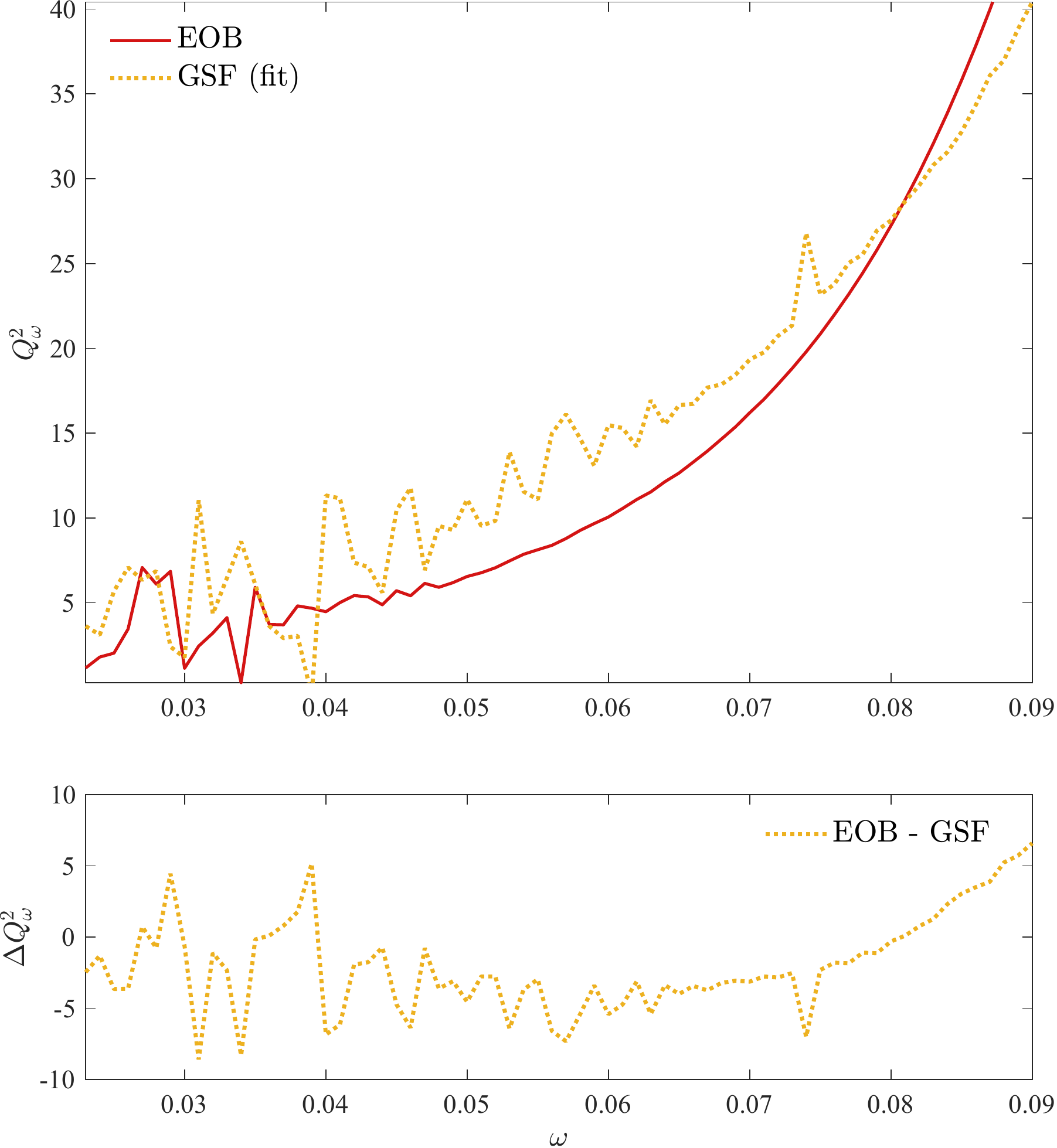} 
\caption{\label{fig:Qis_w500} The coefficients $\Qo^0$ (left), $\Qo^1$ (center) and $\Qo^2$ (right) from the expansion 
\eqref{eq:Qo_exp}, fitted for a set of fixed values of $\omega$ and $\nu$.
The lower panels display the EOB-GSF difference, $\Delta \Qo^i \equiv \Qo^{i, \rm EOB} - \Qo^{i, \rm GSF}  \, (i = 0,1,2)$. }
\end{figure*}

\section{EOB-GSF phasing comparisons}
\label{sec:phasing}
The \PAT{} model was introduced in Ref.~\cite{Wardell:2021fyy} and extensively discussed in
Paper~I, to which we refer the reader for additional technical details. In this section we 
consider mass ratios $q = \left(26, 32, 36, 50, 64, 128, 500\right)$ and compare EOB and
GSF waveform phasings using either time-domain or frequency-domain analyses,
analogously to what was done in Paper~I.

\subsection{Time-domain alignment}
We align waveforms in the time domain via minimization of the 
phase difference on a certain interval as in Paper I (for details see Sec.~VA therein),
and evaluate the phase difference at the time corresponding to the GSF breakdown frequency 
as defined in Eq.~(30) of Paper I.
The results are displayed in Table~\ref{tab:dphis}, while a representative waveform, for $q = 500$, is shown in 
Fig.~\ref{fig:phasing500}. The final dephasings are all positive, which means that the EOB plunge is
in advance of the GSF one, and the EOB evolution is overall faster (less adiabatic) than the GSF one. 
In fact, when comparing these results to Paper I (Table II therein), we can see that the final phase difference
for $q = \left(32, 64, 128\right)$ for \TEOBResumS{} is negative, corresponding to a delayed plunge for 
these mass ratios. The implementation of the GSF-tuned potential and of the hybrid $3^{+19}$PN flux are 
thus effective in allowing for a less adiabatic EOB evolution for large mass ratios.

\subsection{Gauge-invariant analysis}

Again we perform the same gauge-invariant analysis of Paper I. We exploit the adiabaticity parameter
\be
\Qo = \frac{\omega^2}{\dot{\omega}},
\ee
where $\omega \equiv \omega_{22}$ is the $\ell = m = 2$ waveform frequency.
Within the GSF approach, for a fixed value of $\omega$, $\Qo$ can be given as an expansion in $\nu$, i.e.
\be
\label{eq:Qo_exp}
\Qo(\omega; \nu) = \frac{\Qo^{0}(\omega)}{\nu} + \Qo^1(\omega) + \Qo^2(\omega) \nu + O(\nu^2) .
\ee
For details on the different $n$PA contributions $\Qo^{n}$, see Sec.~VI in Paper I. Following common nomenclature in the self-force literature, we refer to quantities that make order-$\nu^{n-1}$ contributions to the orbital phase as ``$n$th post-adiabatic order'' ($n$PA).

\begin{figure}[t]
\includegraphics[width=0.45\textwidth]{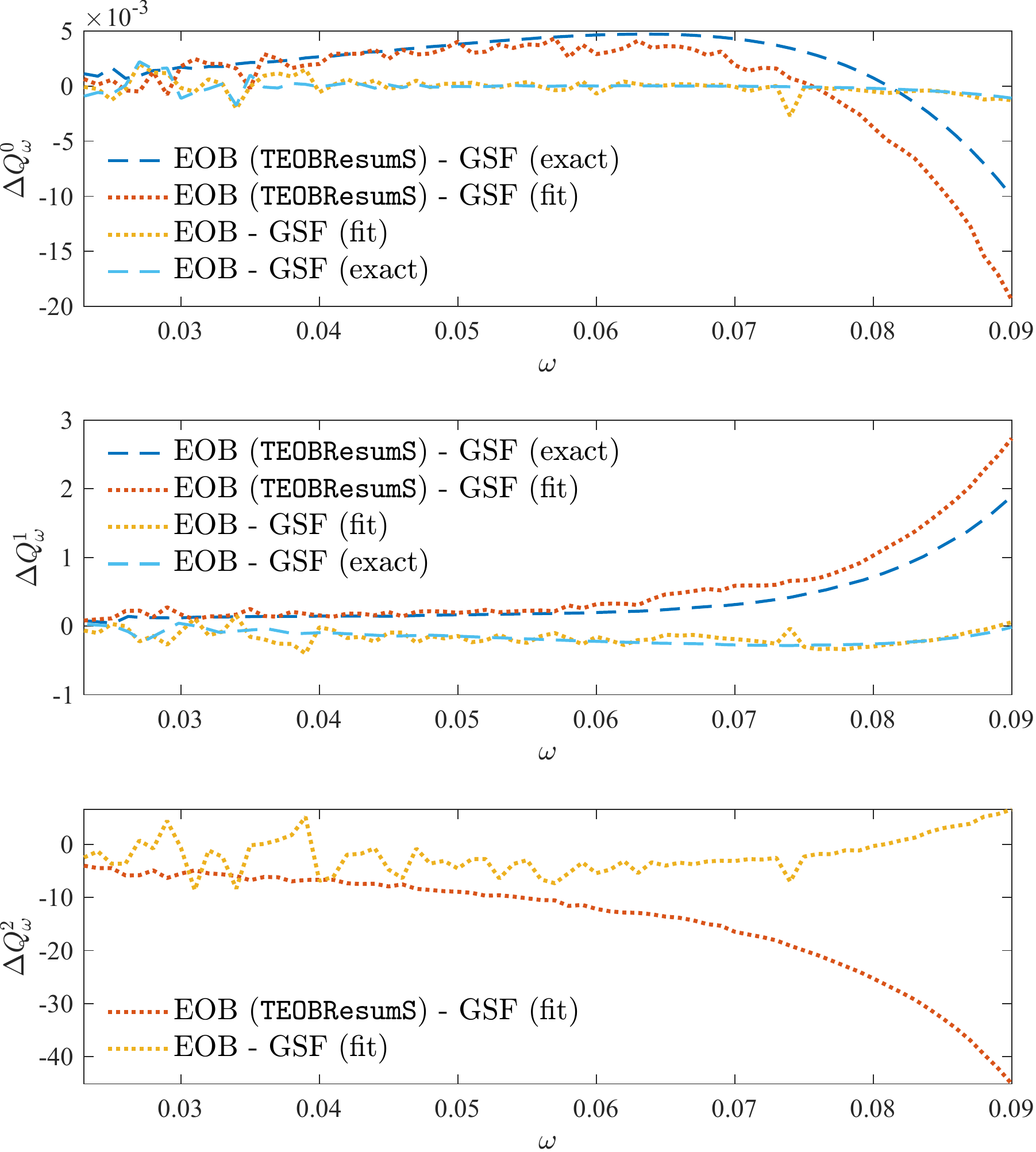} 
\caption{\label{fig:diffQis} Comparing the EOB/GSF differences in the coefficients $\Qo^0$, $\Qo^1$ and $\Qo^2$ 
found in this paper with respect to Paper I, whose results are labeled as \TEOBResumS{}. }
\end{figure}

Given the resummed structure  of the EOB Hamiltonian, the actual $\Qo^{\rm EOB}$ has in fact 
an {\it infinite} number of $\nu$-dependent terms and Eq.~\eqref{eq:Qo_exp} is formally obtained by expanding in $\nu$.
\PAT{} likewise contains an infinite number of terms in Eq.~\eqref{eq:Qo_exp}, but it only yields complete information about the first two terms, $\Qo^0$ and $\Qo^1$; higher-order GSF calculations will lead to different results for the higher-order coefficients $\Qo^n$ with $n>1$.
Our aim here is to extract the three functions $\Qo^0$, $\Qo^1$ and $\Qo^2$ from \PAT{}
and \TEOBResumS{} and compare them. This will give us a precise quantitative understanding 
of the differences between the two models in the limit of small $\nu$. For the fit we use the same procedure described in Paper I,
using mass ratios $q = {26, 32, 36, 50, 64, 128, 500}$, and a range $[\omega_{\rm min},\omega_{\rm max}] = [0.023, 0.09]$ 
with spacing $\Delta \omega = 0.001$.
For each value of $\omega$ we fit $Q_\omega(\omega;\nu)$ using Eq.~\eqref{eq:Qo_exp}. 
The obtained coefficients are plotted in Fig.~\ref{fig:Qis_w500}, along with the exact GSF results 
for $\Qo^0$ and $\Qo^1$ (as computed in Paper I). All three contributions to the $\Qo$ expansion now show a good 
EOB/GSF agreement, especially concerning $\Qo^0$ and $\Qo^1$. In Fig.~\ref{fig:diffQis} we compare these results to those 
of Paper I. First, we see how the results concerning the GSF fit and the exact $\Qo$ are 
more consistent here with respect to Paper I. This is due to the different choice for the mass ratios included in the fit,
as already pointed out in Paper I (see Fig.~11 therein). Then from Fig.~\ref{fig:diffQis} we can 
infer that the new hybrid $3^{+19}$PN flux draws the EOB $\Qo^0$ nearer to the GSF one,
while the GSF-tuned contribution to the EOB potential is responsible for the enhancement in $\Qo^1$.
A deeper justification for this will be given in the following, considering analytical expressions for the three coefficients.

To assess how much each term in the expansion of $\Qo$ impacts the phasing, 
we can estimate three contributions  to the phase difference on the frequency interval $(\omega_1,\omega_2):$
\begin{align}
\label{eq:dphis}
\Delta \phi_0 &\equiv \frac{1}{\nu}  \int_{\omega_1}^{\omega_2}\log(\omega) \; \left( \Qo^{0, \rm EOB} -  \Qo^{0, \rm GSF} \right),\\
\Delta \phi_1 &\equiv \int_{\omega_1}^{\omega_2} \log(\omega) \; \left( \Qo^{1, \rm EOB} -  \Qo^{1, \rm GSF}\right), \\
\Delta \phi_2 &\equiv \nu \int_{\omega_1}^{\omega_2} \log(\omega) \; \left( \Qo^{2, \rm EOB} -  \Qo^{2, \rm GSF}\right),
\end{align}
so that the total phase difference between $(\omega_1,\omega_2)$ is
\be
\Delta\phi^{\rm EOBGSF}_{(\omega_1,\omega_2)}=\Delta\phi_0 + \Delta\phi_1 + \Delta\phi_2.
\ee
The result of this calculation over the frequency interval $(\omega_1,\omega_2)=(0.023,0.09)$ is displayed 
in Table~\ref{tab:dphi_Qis}. As already stressed in Paper I, the results of the integration of $\Qo$
on a given frequency interval cannot be compared to the phase differences obtained via time-domain
alignment of the waveforms. The phase differences here are all negative due to
the fact that the GSF evolution is more adiabatic than the EOB one for these mass ratios and in this frequency range
(compare with Table~IV in Paper I). We also see that the absolute value is decreasing as the mass ratio increases, correspondingly
to the fact that $\Delta \phi_0$ becomes progressively more dominant, while the inverse is true for $\Delta \phi_2$;
when it comes to higher mass ratios the EOB/GSF consistency in $\Qo^0$ is more important than their disagreement in $\Qo^2$.

\begin{table}[t]
\begin{center}
\begin{ruledtabular}
\begin{tabular}{c c | c c c}
$q$ & $\omega_{22}^{\rm GSF_{break}}$  
       & $[\omega_L, \omega_R]$ 
      & $\Delta\phi_{\rm end}^{\rm EOBGSF}$  \\
\hline
\hline
26 &  0.1135 & [0.023, 0.026] & $0.42151$  \\ 
32 &  0.1146 & [0.023, 0.027]  & $0.36474$ \\ 
36 & 0.1151 &  [0.023, 0.027] & $0.33283$ \\
50 &0.1167  &  [0.023, 0.027]  & $0.29113$ \\
64 &0.1178  &  [0.023, 0.028]& $0.25885$  \\
128 & 0.1207  &  [0.023, 0.027]  & $0.23042$ \\
500 & 0.1251 & [0.023, 0.027] & $0.21041$ \\
\end{tabular}
\end{ruledtabular}
\end{center} 
\caption{From left to right, the columns report: the mass ratio $q$;
the GSF breakdown frequency, $\omega_{22}^{\rm GSF_{break}}$ as defined in Eq.~(30) of Paper~I; 
the alignment interval used in the time-domain phasing;
the phase difference, computed up to $\omega_{22}^{\rm GSF_{break}}$, 
using the time-domain alignment.}
\label{tab:dphis}
\end{table}

\begin{table}[t]
\begin{center}
\begin{ruledtabular}
\begin{tabular}{c c c c c}
$q$ &  $\Delta \phi_0$ & $\Delta \phi_1$ & $\Delta \phi_2$ & $\Delta \phi_{[\omega_1, \omega_2]}$\\
\hline
\hline
26 & 0.0011045 & $-0.20971$ & $-0.10711$ & $-0.31572$ \\
32 & 0.0013406 & $-0.20971$ & $-0.088252$ & $-0.29662$ \\
36 & 0.001498 & $-0.20971$ & $-0.078977$ & $-0.28719$ \\
50 & 0.0020493 & $-0.20971$ & $-0.057734$ & $-0.26539$ \\
64 & 0.0026006 & $-0.20971$ & $-0.045494$ & $-0.2526$ \\
128 & 0.0051215 & $-0.20971$ & $-0.023101$ & $-0.22769$ \\
500 & 0.019776 & $-0.20971$ & $-0.0059827$ & $-0.19592$ \\
\end{tabular}
\end{ruledtabular}
\end{center} 
\caption{From left to right, the columns report: the mass ratio $q$, the phase differences due to the first three term in
the expansion of $\Qo$, and the sum of these latter. The $\Delta \phi$'s are obtained using the definition~\eqref{eq:dphis},
integrating on the frequency interval $[\omega_1, \omega_2] = [0.023, 0.09]$.
}
\label{tab:dphi_Qis}
\end{table}

\section{Understanding the $\Qo$ expansion}
\label{Qomg:anlyt}
The behavior of $\Qo$ and of its three different contributions, $(\Qo^0, \Qo^1, \Qo^2)$,
can be understood analytically when working in the circular approximation. Assuming for simplicity
that the gravitational wave frequency is $\omega_{22}=2\Omega$, where $\Omega$ is the
EOB orbital frequency\footnote{We are here neglecting in the EOB waveform the additional 
contributions to the frequency that come from the resummed tail factor and from the residual phase 
correction $\delta_{22}$~\cite{Damour:2008gu}. Note that the contribution to the tail cannot
be extracted analytically in closed form; see Appendix~E.2 of Ref.\cite{Nagar:2018zoe}. 
Nonetheless, this approximation does not change the conclusions of our reasoning below.}, 
we have
\be
\Qo =  2\frac{\Omega^2}{\dot{\Omega}} \, ,
\ee
and using the frequency parameter  $x \equiv \Omega^{2/3}$, we have
\begin{align}
\dot{\Omega} &= \de_j \Omega \de_t j = \frac{3}{2} x^{1/2} \frac{\de x}{\de j} \hat{\mathcal{F}}_\ph \, ,
\end{align}
where $j \equiv J^{\rm circ} / \mu M$ is the orbital angular momentum along circular orbits
and we replaced $\de_t j =\hat{\mathcal{F}_\ph}$. The angular momentum flux is written as
$\hat{\mathcal{F}_\ph} = \mathcal{F}_\ph / \nu =\nu \mathcal{F}_\ph^{\rm 1SF} + \nu^2 \mathcal{F}_\ph^{\rm 2SF} + \nu^3{\cal F}_\varphi^{\rm 3SF}$
(i.e., as a 2PA expansion) to meaningfully compare EOB and GSF contributions.
Note, however, that the complete EOB flux, which is summed up to $\ell=8$, has many more 
$\nu$-dependent terms because it incorporates all the known $\nu$ dependence\footnote{The precise evaluation
of the exact $\nu$ order is tricky because of the resummed nature of the EOB fluxes. However, if we focus
only on the $\nu$ dependence of the leading, Newtonian, prefactor of each mode, $F_\lm^{\rm Newt}$, we see
that the EOB flux is at least partly $\mathcal{F}_\ph^{\rm 9SF}$.} up to 3PN order. 
The $\Qo$ function can be rewritten as
\begin{align}
\label{Qo_exp}
\Qo(x) &=\frac{4}{3} \frac{x^{5/2}}{\nu\FISF} \left\{1 - \nu \frac{\FIISF}{\FISF} \right .\nonumber\\
           &\left.- \nu^2 \left[ \frac{{\cal F}^{\rm 3SF}_\varphi}{\FISF} - \left(\dfrac{\FIISF}{\FISF}\right)^2\right]\right\} {\de_x j} .
\end{align}
Within the EOB approach, the angular momentum along circular orbits 
is given by
\be
\label{eq:p_circ}
j^2=-\frac{\partial_u A}{\partial_u(u^2A)}=-\frac{\partial_u A}{2u \tilde A}\,,
\ee
where $u = M/r$ is the dimensionless gravitational Newtonian potential
and $\tilde A(u; \nu) \equiv A(u;\nu)+\frac12 u \partial_u A(u;\nu)$.
Considering the interbody EOB potential $A$ as a formal expansion up to $\nu^2$,
\be
A(u;\nu) =1-2u +\nu a_{1}(u)+\nu^2 a_{2}(u) +O(\nu^3) \ ,
\ee
from Eq.~\eqref{eq:p_circ} one obtains $j$ at 2PA order as
\begin{align}
\label{eq:ju}
j(u) = &-\frac{1}{32 (1-3 u)^2} \sqrt{\frac{1}{u-3 u^2}} \Big[ \left(-32 +192 u -288 u^2\right) \nonumber \\
	&+ \nu \left( a_1'(u) (8 -40 u + 48 u^2) + a_1(u) (16 -48 u) \right)  \nonumber  \\
	&+ \nu^2 \Big( a_1'(u)^2 (1 -8 u + 12 u^2) -4 a_1(u) a_1'(u)  \nonumber \\
	&-12 a_1(u)^2 + a_2'(u) (8 -40 u + 48 u^2) \nonumber \\ 
	& +16 (1-3 u) a_2(u) \Big)\Big]  + O(\nu^3).
\end{align}
To obtain $j(x)$ to complete the expression of $\Qo(x)$ in Eq.~\eqref{Qo_exp},
we take advantage of Eq.~(2.21) of Ref.~\cite{Bini:2016cje}, which gives $u(x)$.
This relation reads
\begin{align}
\label{eq:ux}
u&=x -\nu U_1(x; a_1'(x)) \nonumber\\
  &+\nu^2 V_2(x;a_1(x),a_1'(x),a_2'(x))+O(\nu^2)\,,
\end{align}
where 
\begin{align}
&U_1(x; a_1'(x))
\!= \! -\frac16 x \left[ a_1'(x)-4 \left(1-\frac{1-2x}{\sqrt{1-3x}}  \right)\right], \  \\
 &U_2(x;a_1(x),a_1'(x),a_2'(x)) \! = \!  -\frac13 \frac{x (1-4 x)}{ (1-3 x)^{3/2}} a_1(x)\nonumber\\
&-\frac16  x a_2'(x)-\frac{1}{36}x [a_1'(x)]^2 \nonumber\\
\nonumber\\
&+\left(\frac{ x (1-2 x) (2-3 x)}{18(1-3 x)^{3/2}}-\frac19  x\right)a_1'(x)\nonumber\\
&-\frac{16 x (1-2 x)}{9 (1-3 x)^{1/2}}+\frac{8x(2-7 x+4 x^2)}{9(1-3 x)} \ ,\\
&V_2(x;a_1(x),a_1'(x),a_2'(x))
 \equiv U_1(x; a_1'(x)) \left(\frac{d}{dx}U_1(x; a_1'(x))\right)\nonumber\\
&-U_2(x;a_1(x),a_1'(x),a_2'(x))\,.
\end{align}
By combining Eq.~\eqref{eq:ju} and Eq.~\eqref{eq:ux} we have $j(x)$, and
we can finally evaluate explicitly Eq.~\eqref{Qo_exp} as a 
function of $(\FISF, \FIISF,\FIIISF,a_1(x), a_2(x))$, to obtain
\be\label{Q0 exact}
\Qo^0 = - \frac{2}{3} \frac{(1- 6x)}{(1 - 3x)^{3/2}} \frac{x^{3/2}}{\FISF} \ ,
\ee
\begin{widetext}
\begin{align}
\Qo^1 &= -\frac{x }{9 (\FISF)^2 (1-3 x)^3} \Bigg\{ \FISF \Bigg[\sqrt{1-3 x} \bigg(\left(-54 x^2+24 x-2\right)
   a_1'(x)+\left(36 x^3-24 x^2+4 x\right) a_1''(x)   \nonumber\\
   &+  (36 x-3) a_1(x)+72 x^2-12 x+2\bigg)+72 x^3-72 x^2+14 x-2\Bigg]
   + x \FIISF \sqrt{1-3 x} \left(-108 x^2+54 x-6\right)\Bigg\} \ , \label{Q1 exact}\\
\Qo^2 = &  \frac{x }{108 (\FISF)^3 (1-3 x)^{5}}
   \Bigg\{ (\FISF)^2 \Bigg[ \sqrt{1-3 x} \Bigg(a_1(x) \big(\left(2268 x^3-1620 x^2+324 x-12\right)a_1'(x) \nonumber\\
   &+\left(-648 x^4+648 x^3-216 x^2+24 x\right) a_1''(x)-5184 x^3+1944 x^2-108 x+12\big) \nonumber\\
   &+a_1'(x) \big(\left(-324 x^5+648 x^4-432 x^3+120 x^2-12 x\right)a_1''(x) \nonumber\\
   &+\left(-648 x^6+864 x^5-432 x^4+96 x^3-8 x^2\right) a_1^{(3)}(x)+6480 x^4-4752x^3+1080 x^2-96 x+8\big) \nonumber\\
   &+\left(-486 x^4+351 x^3-54 x^2-9 x+2\right) a_1'(x)^2+\left(-648x^6+864 x^5-432 x^4+96 x^3-8 x^2\right) a_1''(x)^2 \nonumber\\
   &+\left(-3888 x^5+3888 x^4-1296 x^3+144 x^2\right) a_1''(x)+\left(2592 x^6-3456 x^5+1728 x^4-384 x^3+32 x^2\right)a_1^{(3)}(x) \nonumber\\
   &+\left(-1458 x^2+567 x-27\right) a_1(x)^2+\left(5832 x^4-6480 x^3+2592 x^2-432x+24\right) a_2'(x) \nonumber\\
   &+\left(-3888 x^5+5184 x^4-2592 x^3+576 x^2-48 x\right) a_2''(x)+\left(-3888 x^3+2916 x^2-648 x+36\right) a_2(x) \nonumber\\ 
   &-14976 x^4+11424 x^3-3528 x^2+676 x-56\Bigg)+\left(2592 x^5-6480 x^4+3816 x^3-840 x^2+88 x-8\right) a_1'(x) \nonumber\\
   &+\left(-2592 x^6+5616 x^5-3888 x^4+1104 x^3-112 x^2\right) a_1''(x) \nonumber\\
   &+ \left(5184 x^7-9504 x^6+6912 x^5-2496 x^4+448 x^3-32 x^2\right) a_1^{(3)}(x) \nonumber\\
   &+\left(2160 x^3-576 x^2-48 x\right) a_1(x)-18144 x^5+30240 x^4-16200 x^3+4488 x^2-760 x+56\Bigg] \nonumber\\
   &+\FISF \Bigg[\FIISF \Bigg(\sqrt{1-3 x} \big(\left(-5832 x^4+6480 x^3-2592 x^2+432 x-24\right) a_1'(x) \nonumber\\
   &+\left(3888 x^5-5184 x^4+2592 x^3-576 x^2+48 x\right) a_1''(x)+\left(3888 x^3-2916 x^2+648 x-36\right) a_1(x) \nonumber\\
   &+7776 x^4-6480 x^3+1944 x^2-288 x+24\big)+7776 x^5-12960 x^4+7560 x^3-2088 x^2+312 x-24\Bigg) \nonumber\\
   &+\FIIISF \sqrt{1-3 x} \left(11664 x^4-13608 x^3+5832 x^2-1080 x+72\right)\Bigg] \nonumber\\ 
   &+(\FIISF)^2 \sqrt{1-3 x} \left(-11664 x^4+13608 x^3-5832 x^2+1080 x-72\right) \Bigg\} \ .
\end{align}
\end{widetext}

As one should expect, the 0PA term~\eqref{Q0 exact} is identical to the formula derived within GSF theory; 
see Eq.~(B6) in Paper I [with (12), (14), and the relation ${\cal F}^{(1)}=-\Omega\FISF$, where ${\cal F}^{(1)}$ 
is the leading-order flux of energy to infinity and into the black hole]. The 1PA term~\eqref{Q1 exact} can similarly 
be compared to the first term in Eq.~(B7) [with (13)] of Paper I if we note that $a_1(x)$ is directly related 
to the binding energy $\hat E_{\rm SF}$ appearing there~\cite{Akcay:2012ea}.

We are now in the position of understanding in detail the results of Fig.~\ref{fig:Qis_w500}.
First, since only  $\FISF$ contributes to $\Qo^0$, the excellent EOB/GSF agreement in this coefficient
we obtain here is mostly due to the inclusion of the $3^{+19}$PN flux at infinity\footnote{Note that both the 
\PAT{} and EOB model implement a contribution to the flux due to absorption by the horizon of the two black holes. 
This is different in the two models, but the differences do not seem to be important up to $q=500$. Effects due to 
horizon absorption are discussed in Sec.~\ref{sec:horizon}.}. 
On the other hand, $\Qo^1$ is function of $\FISF$, $a_1$ and $\FIISF$. The good EOB/GSF consistency of $\Qo^1$
suggests that the accurate modelling of $\FISF$ and $a_1$ (as is the case because this function is GSF-informed) 
is more important than the modelling of $\FIISF$ (that is different in the two models)  to correctly capture this contribution.
Finally, we see that $\Qo^2$ depends on $a_2$, which is zero in both models, on $\FIISF$ and on $\FIIISF$, that is 
zero in \PAT{}, but nonzero in EOB. The $\Qo^2$ differences should then mostly come from  $\FIISF$ and $\FIIISF$. 

The availability of the analytic $(\Qo^0,\Qo^1,\Qo^2)$ allows us to devise a more precise
interpretation of the analogous analysis shown in Fig.~10 of Paper~I. Contrasting with what we obtain
here, one has to keep in mind that: (i) for $\Qo^0$, most of the EOB/GSF difference obtained 
considering the standard \TEOBResumS{} is indeed due to the use of a flux that does not include the same amount of 
test-mass information used here, as already pointed out in Paper~I; (ii) for what concerns $\Qo^1$, Paper~I
uses a different, non-GSF-informed, but NR-informed, expression for $a_1$, which explains why the disagreement
in $\Qo^1$ is larger than the one shown here; (iii) in addition, in Paper~I we also had information {\it beyond}
$a_1$, related to $a_2$ and higher (effective) terms informed by NR simulations, which similarly explains 
the larger disagreement found there in $\Qo^2$. This intuitive understanding can be made more quantitative as follows.
First, let us recall that the expression for $A$ used in \TEOBResumS{} stems from
the formal 5PN expression
\begin{align}
\label{eq:A5pn}
A_{\rm 5PN}(u) = & \;  1 -2 u + 2 \nu u^3 + \left( \frac{94}{3} - \frac{41}{32} \pi^2 \right)\nu u^4 \nonumber \\ 
			   & + \nu [ a_5^c(\nu) + a_5^{\ln} \ln u ] u^5 \nonumber \\ 
			   & + \nu [ a_6^c(\nu) + a_6^{\ln}(\nu) \ln u ] u^6  ,
\end{align}
where $a_6^c$ plays the role of an effective 5PN parameter that is informed by NR simulations
once the expression above is replaced by the Pad\'e resummed potential 
$A(u; a_6^c(\nu); \nu) \equiv P^1_5 [A^{\rm 5PN}(u)]$, where $P^1_5$ indicates
the $(1,5)$ Pad\'e approximant.
\begin{figure}[t]
\includegraphics[width=0.45\textwidth]{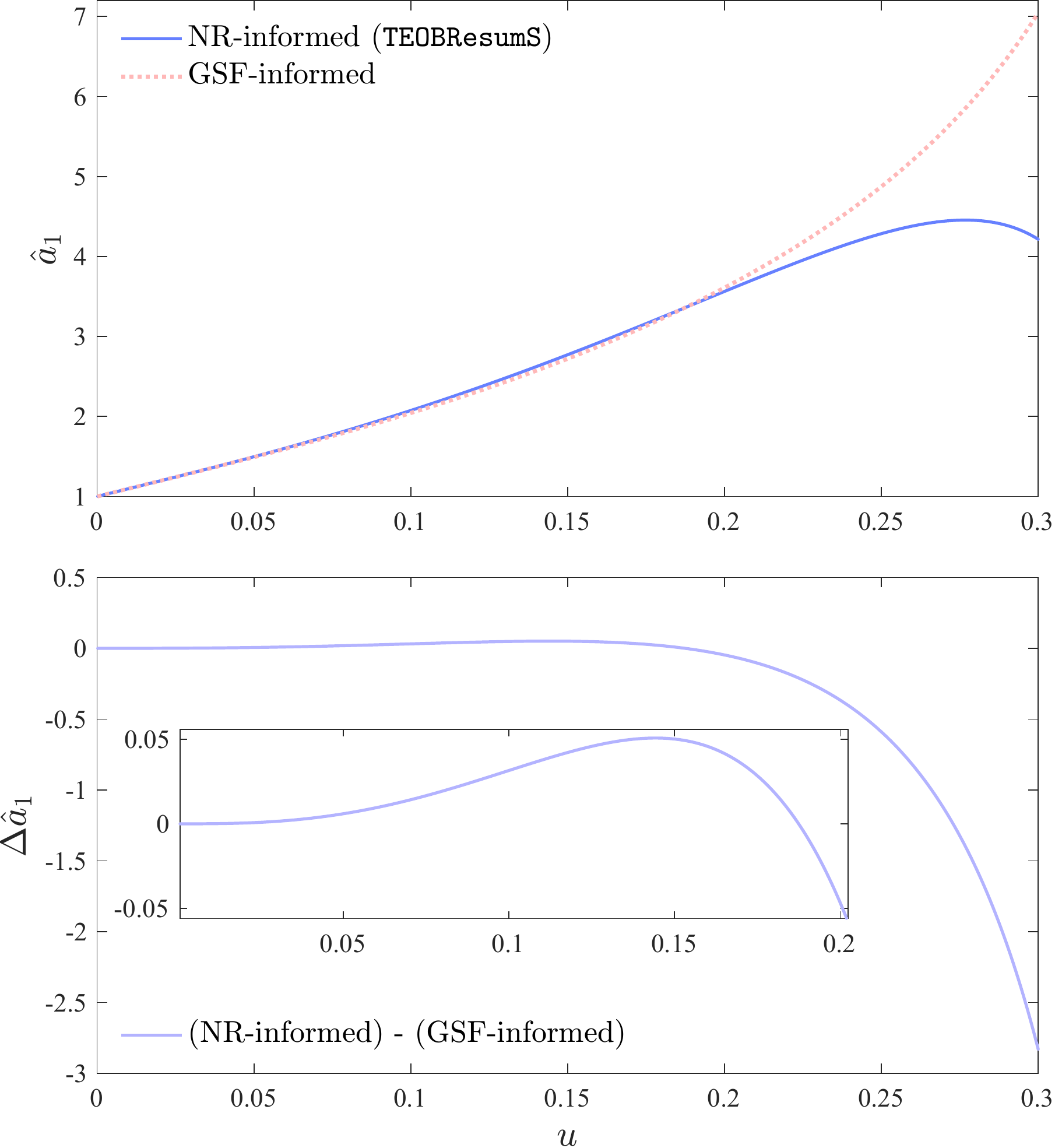} 
\caption{\label{fig:a1} Comparing the GSF-informed $\hat{a}_1$ function defined in Eq.~\eqref{eq:hata} 
to the $\hat{a}_1$ coming from the expansion in $\nu$ 
of the NR-informed Pad\'e-resummed $A$ potential entering the standard \TEOBResumS{}.}
\end{figure} 
In Eq.~\eqref{eq:A5pn} 2PA terms explicitly appear as
\begin{align}
a_5^{\ln}    &= \dfrac{64}{5} \ , \\
a_5^c(\nu) &= a^c_{5,0} + \nu a^c_{5,1} \ ,\\
a_{5,0}^c   &=\dfrac{2275}{512}\pi^2-\dfrac{4237}{60}+\dfrac{28}{5}\gamma_{\rm E}+\dfrac{256}{5}\ln2 \ , \\
a_{5,1}^c   &=\dfrac{41\pi^2}{32}-\dfrac{221}{6} \ , \\
a_6^{\ln}(\nu)   &= -\left(\dfrac{7004}{105}+\dfrac{144}{5}\nu\right) ,
\end{align}
where $\gamma_{\rm E}=0.577216\dots$ is Euler's constant.
Although the coefficient $a_6^c(\nu)$ is analytically known at 3PA modulo one missing 2PA 
coefficient, $a_6^{\nu^2}$~\cite{Bini:2019nra,Bini:2020wpo}, we keep it here 
as an unknown function to be informed by NR simulations. In particular, we use 
the expression of the effective $a_6^c(\nu)$ is given by 
Eqs.~(33)-(38) of Ref.~\cite{Nagar:2020pcj}.
The Pad\'e resummed potential can be expanded in $\nu$ as
\be
A(u; \nu) \approx 1 - 2u + \nu a^{\rm teob}_1(u) + \nu^2 a^{\rm teob}_2(u) + O(\nu^3).
\ee
It is then convenient to normalize $a^{\rm teob}_1$ as 
\be
\label{eq:hata}
\hat{a}^{\rm teob}_1(u) = \frac{a_1^{\rm teob}(u)}{ 2 u^3 E(u)} ,
\ee
where $E(u) = (1 -2 u)/\sqrt{1 - 3u}$ to ease the comparison with 
the GSF-informed function that diverges at $u=1/3$. In Fig.~\ref{fig:a1} 
we compare the NR-informed $\hat{a}_1^{\rm teob}$ and the GSF-informed $\hat{a}_1$ 
used in the EOB model we are considering in this work. The difference is nonnegligible
and accounts quantitatively of (part of) the differences between the EOB and GSF 
$\Qo^1$ found in Paper~I.  Let us also note that the $a_2^{\rm teob}$ function is 
nonzero, and thus provides a clear justification for the large EOB/GSF disagreement 
in $\Qo^2$ that was found in Paper I. 

\begin{figure}[t]
\includegraphics[width=0.45\textwidth]{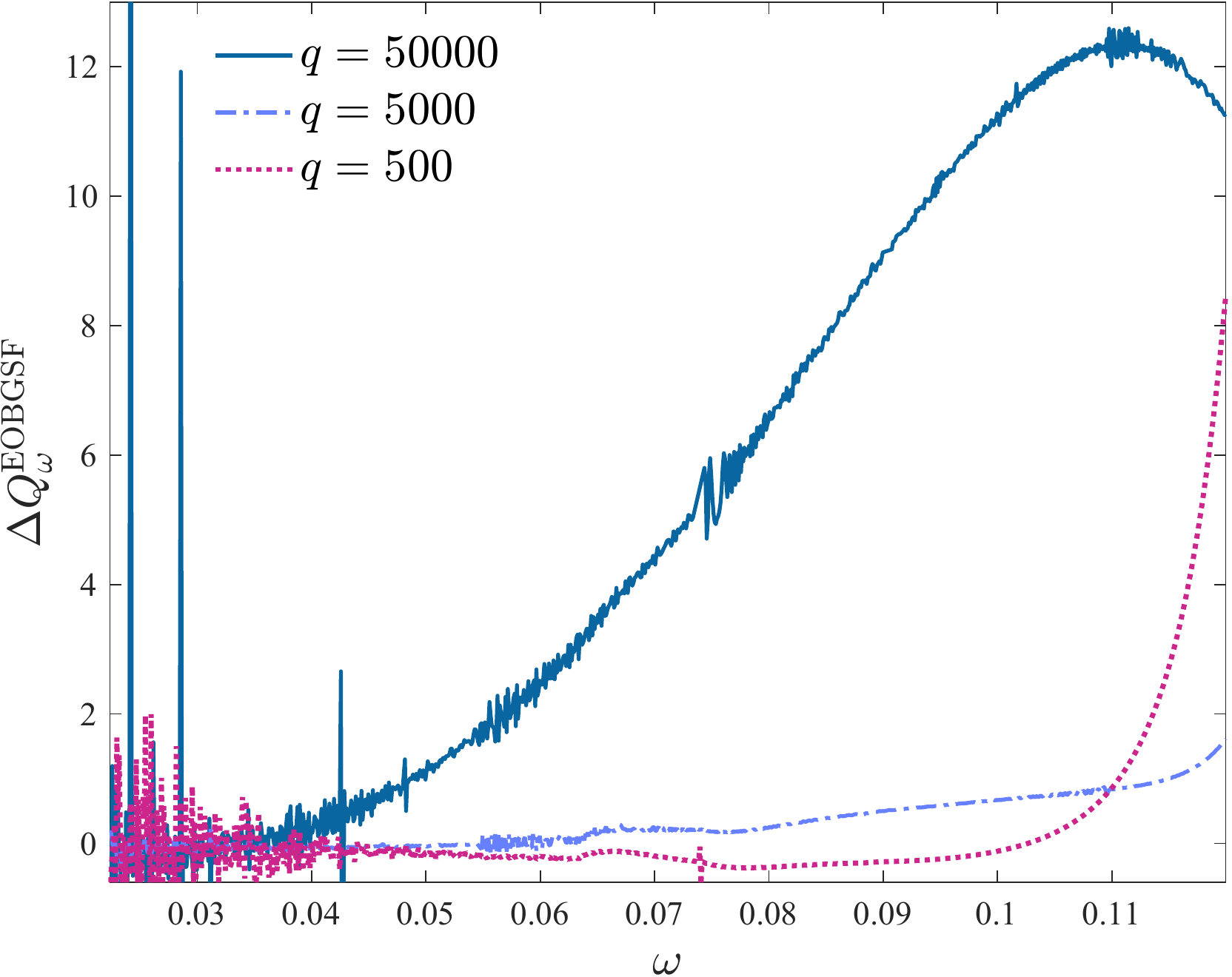} 
\caption{\label{fig:IMRtoEMR}EOB/GSF $\Qo$ difference $\Delta\Qo^{\rm EOBGSF}\equiv \Qo^{\rm EOB}-\Qo^{\rm GSF}$ 
for $q = \left(500, 5000, 50000\right)$ binaries- The integrated phase differences on the frequency range 
$\Delta\omega = (0.0224, 0.12)$ are $\left( 0.07, 0.27, 5.88\right)$~rad respectively. The initial binary separation is $r=20$ 
for each configuration. The EOB/GSF performance progressively worsens as the mass ratio enters into the EMRI regime.}
\end{figure}

\section{Improving the EOB/GSF agreement: the role of the horizon flux}
\label{sec:horizon}
\begin{figure*}[t]
\includegraphics[width=0.32\textwidth]{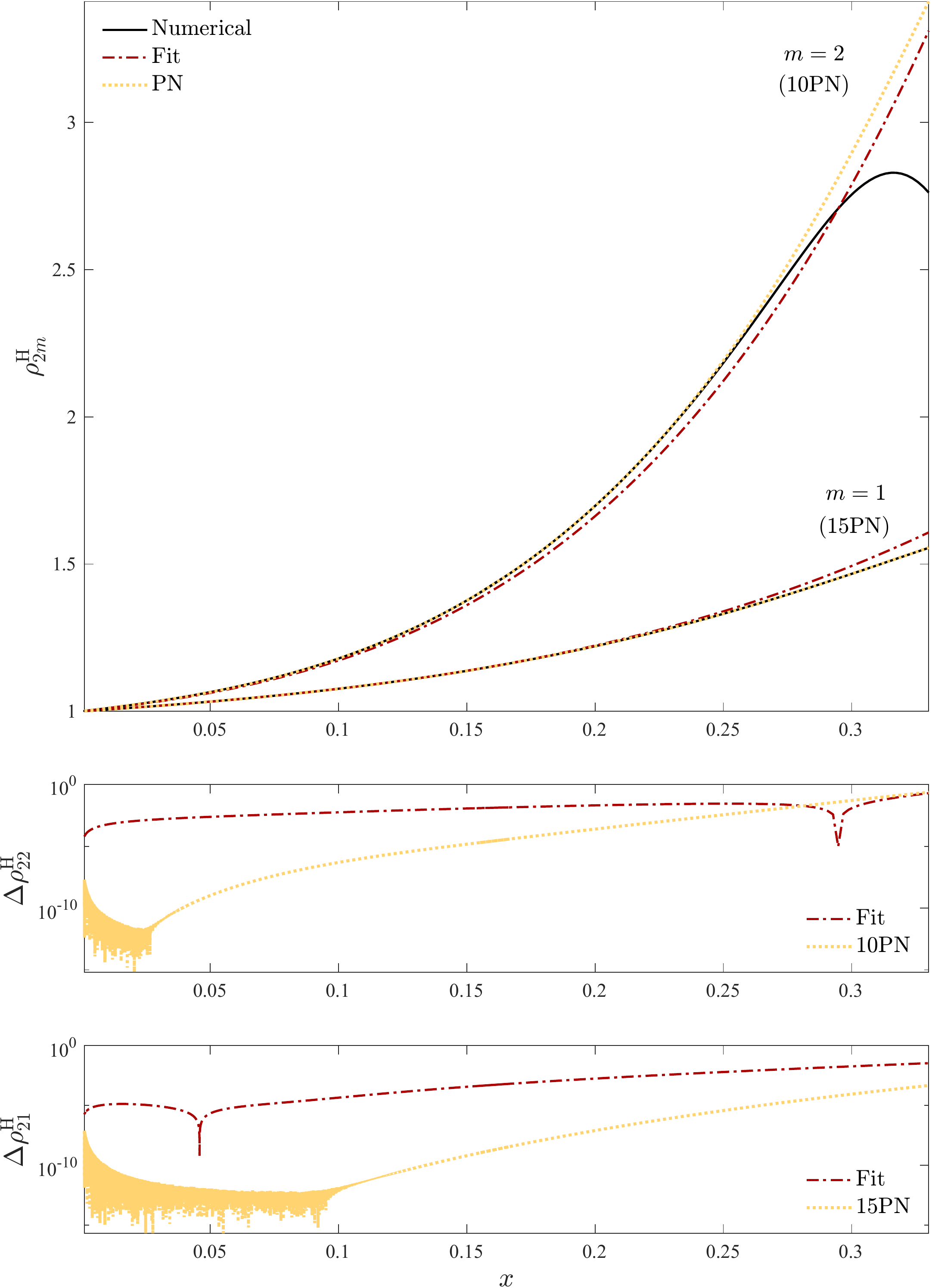} 
\includegraphics[width=0.32\textwidth]{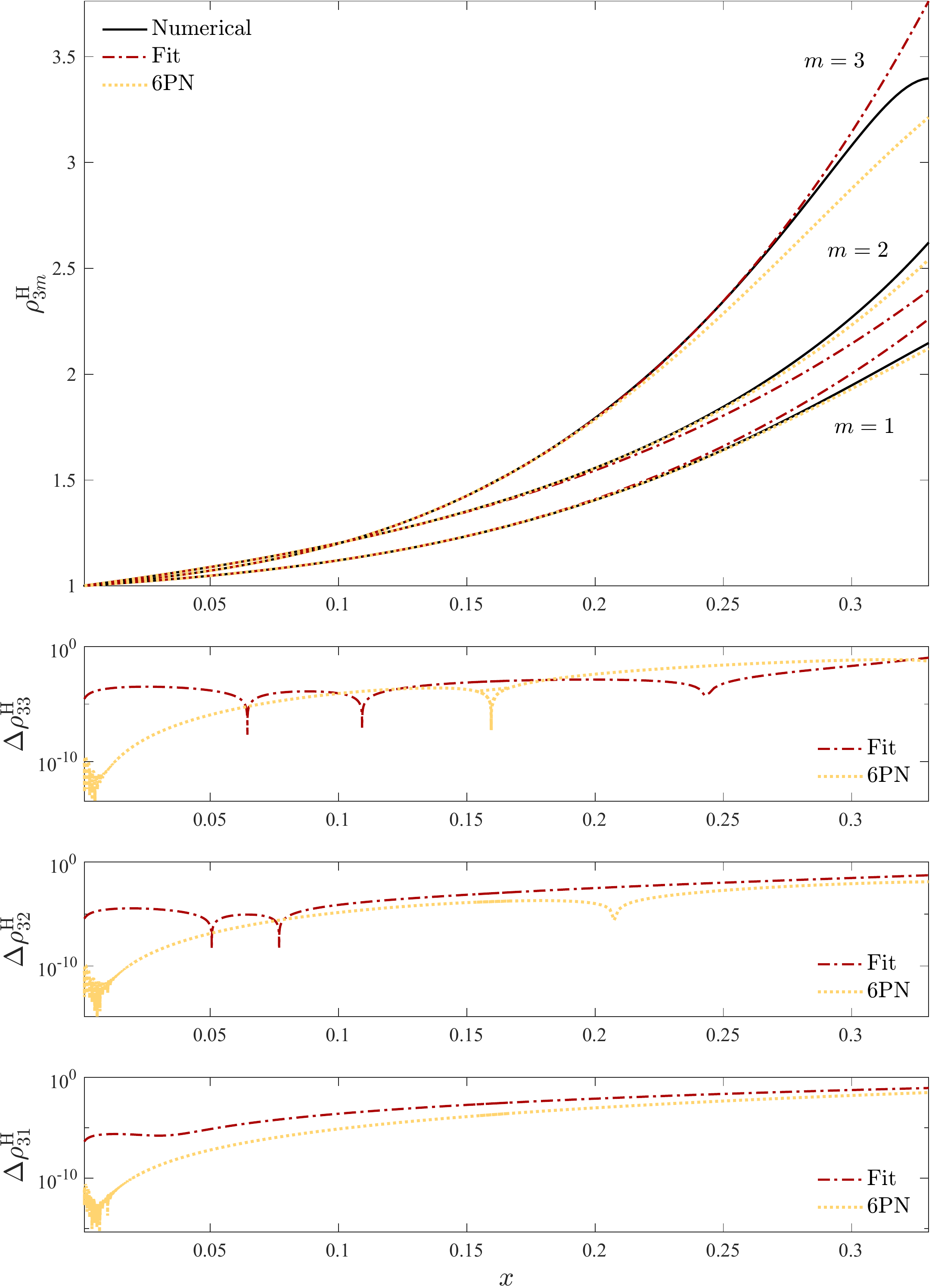} 
\includegraphics[width=0.32\textwidth]{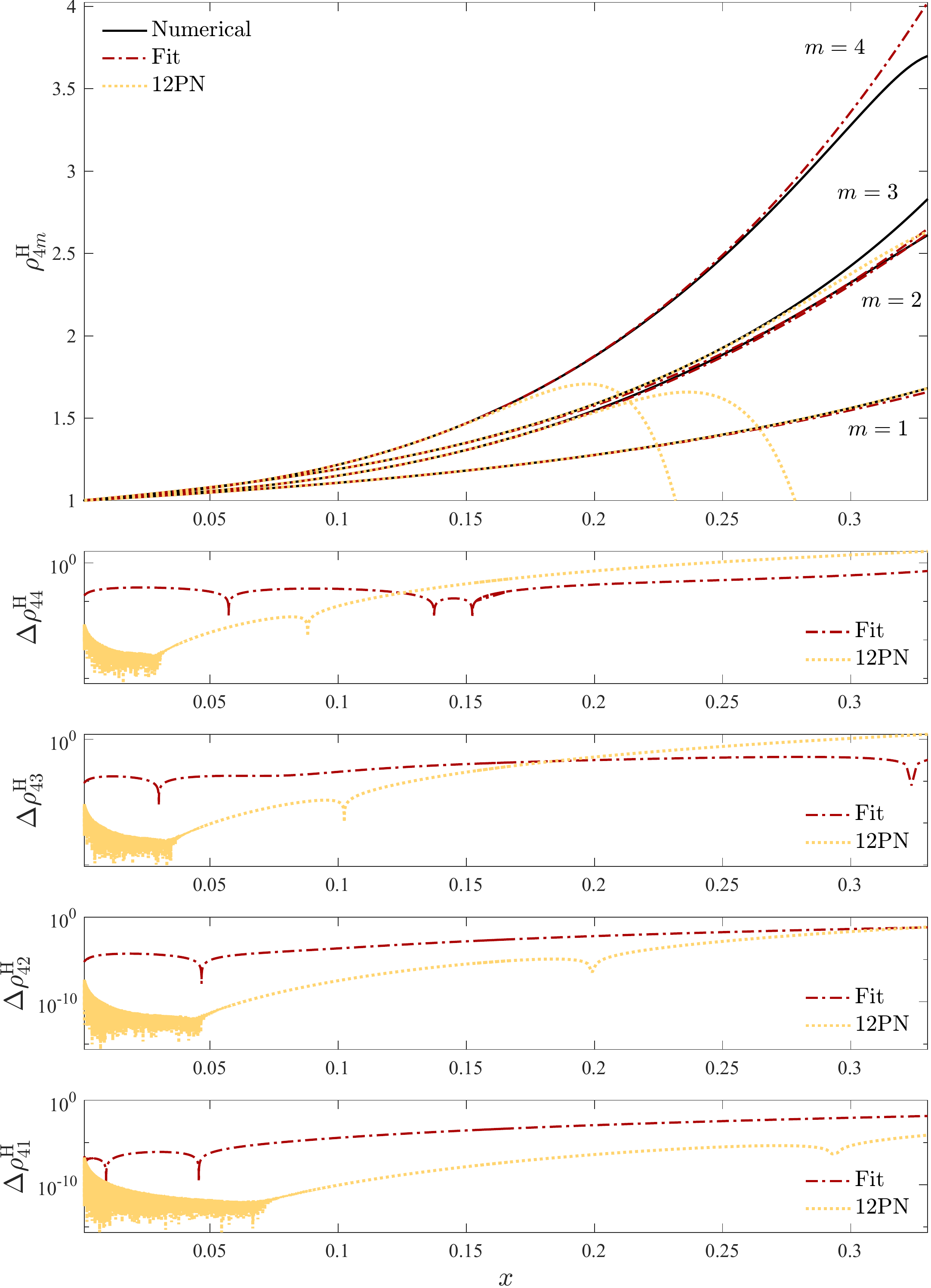} 
\caption{\label{fig:rhoH}Residual amplitude corrections $\rho_\lm^H$ to the horizon flux of a test mass orbiting a Schwarzschild
black hole on circular orbits. The numerical data are compared to the fits of Ref.~\cite{Bernuzzi:2012ku}  and
to the high-PN results of~\cite{Fujita:2014eta}. See text for discussion.}
\end{figure*}
In the previous section we have only focused on mass ratios up to $q = 500$, which 
pertains to the lower boundary of the intermediate-mass-ratio regime. 
Let us now move to considering  even larger mass ratios, so as to span up to the 
extreme-mass-ratio (EMR) regime. Figure~\ref{fig:IMRtoEMR} shows the EOB/GSF $\Qo$ 
differences for $q=\left(500,5000,50000\right)$: it illustrates how the EOB/GSF 
evolution worsens progressively as the mass ratio reaches the EMR regime.
The plot shows that this EOB model is close to reaching a GSF-faithful evolution 
for $q = 5000$, given that $\Delta\Qo^{\rm EOBGSF}\equiv \Qo^{\rm EOB} - \Qo^{\rm GSF}$ 
is of order $1$ at $\omega = 0.12$, but is still far from being sufficiently accurate to model EMRIs. 
However, from the previous $\Qo$ analysis we learnt that as the mass ratio 
increases, the 0PA contribution to the dephasing is more and more relevant. 
Given that $\Qo^0$ only depends on $\FISF$, the inconsistency highlighted 
in Fig.~\ref{fig:IMRtoEMR} should be mostly due to residual differences in $\FISF$
between the EOB dynamics and \PAT{}. This hypothesis is further supported by the fact 
that the disagreement appears to grow linearly with $q$ when moving from $q=5000$ 
to $q=50000$ in Fig.~\ref{fig:IMRtoEMR}, as one would expect from a disagreement in $\Qo^0$.

As briefly mentioned above, 
a difference certainly lies in the modelization of the 1SF horizon flux.
\PAT{} implements the exact 1SF horizon flux summed up to $\ell=30$. 
By contrast, \TEOBResumS{} uses an approximate, though resummed, 
expression that only includes the $\ell=m=2$ and $\ell=2$, $m=1$ multipoles 
as discussed in Refs.~\cite{Bernuzzi:2012ku,Nagar:2011aa}. In particular, the 
PN information beyond the leading order contributions, for each mode, is collected 
into the residual amplitude correction functions, called $\rho_\lm^H$, that are 
the analogous of the $\rho_\lm$'s for the horizon  flux. 
The $\rho_{22}^H$ and $\rho_{21}^H$ functions we are using here are those
introduced and discussed in detail in Sec.~II of Ref.~\cite{Bernuzzi:2012ku}.
They are given by formal 4PN polynomials obtained in the following 
way: (i) first, one was fitting the (multipolar) horizon fluxes of a test-mass
around a Schwarzschild black hole with a rational function and then (ii) this
rational function was expanded up to 4PN order in order to hybridize the 
exact 1PN term with the other three effective terms up to 
4PN order~\cite{Nagar:2011aa}. Reference~\cite{Bernuzzi:2012ku} 
computed corrections to the horizon flux up to $\ell=4$ (see Table~I therein),
but it explicitly considered only the quadrupolar contribution, which was deemed
sufficient for the purposes of that study. By contrast, here we find that
the effect of the higher multipoles is actually nonnegligible and is useful to reduce the 
EOB/GSF gap, as we will discuss next.

\subsection{Improved horizon flux}
\begin{figure}[t]
\includegraphics[width=0.45\textwidth]{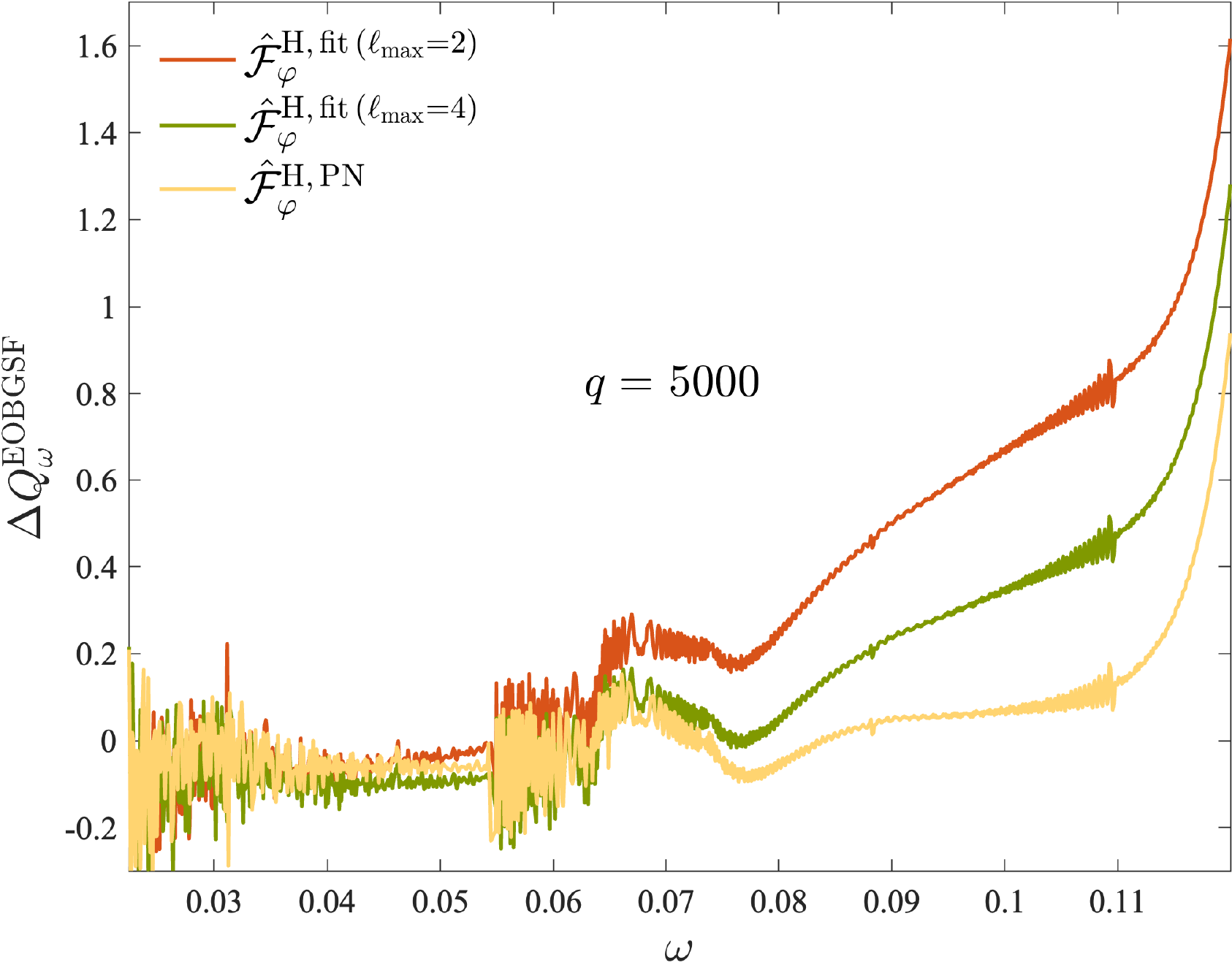} \\
\vspace{2.5mm}
\includegraphics[width=0.45\textwidth]{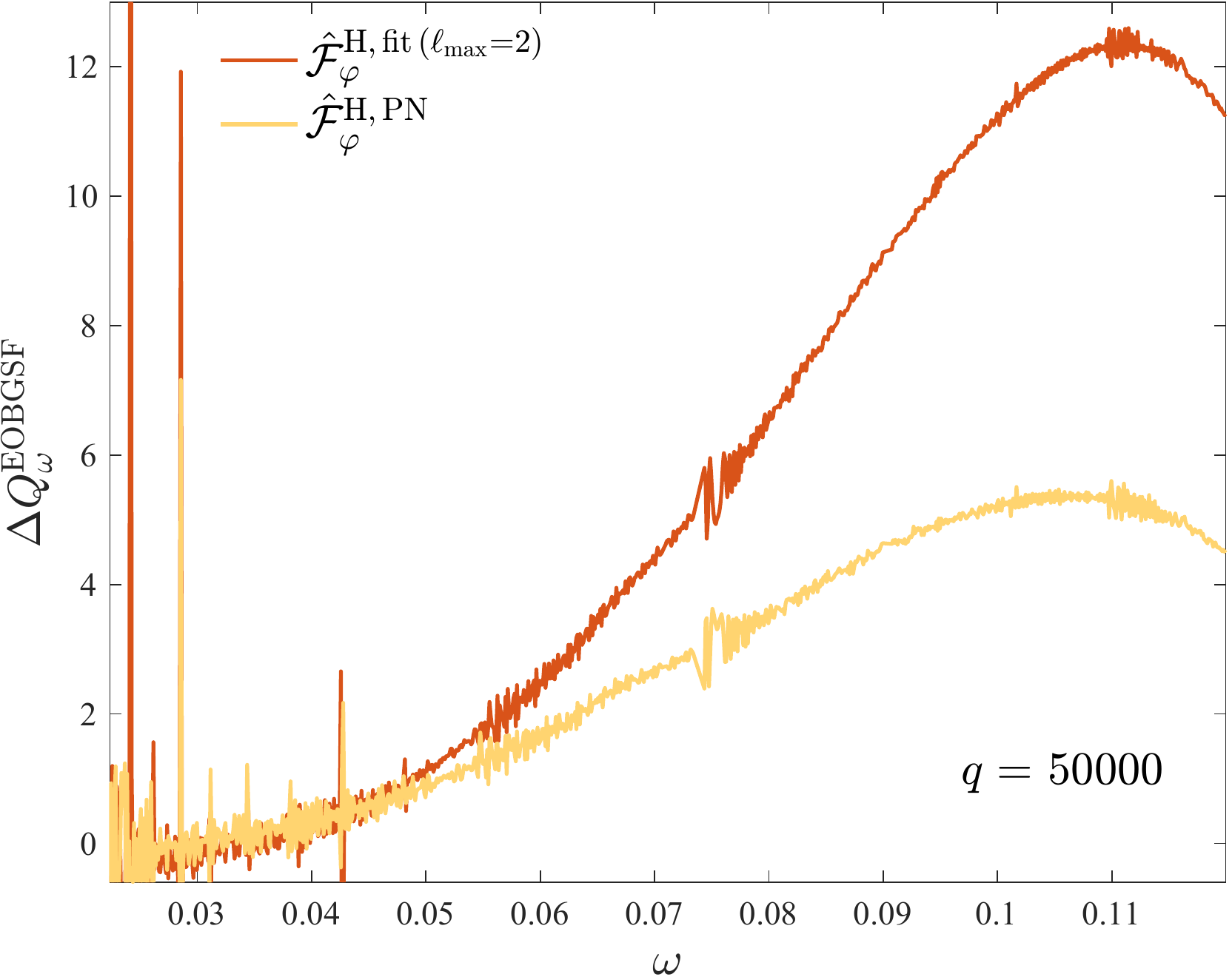} 
\caption{\label{fig:5000_newHF}Impact of the additional terms in the horizon flux on the 
EOB/GSF $\Qo$ agreement for $q = 5000$ and $q=50000$. The initial separation is $r=20$
for each configuration.}
\end{figure}

To understand the impact of horizon absorption on our current results,
let us first remember the structure of the $\rho_\lm^H$ functions, 
up to $\ell=4$, and of their approximations, according to~\cite{Nagar:2011aa}.
For each multipole $(\ell,m)$, Fig.~\ref{fig:rhoH} shows: (i) the exact (numerical)
curves for a test-mass on circular orbits around a Schwarzschild black hole,
as computed in~\cite{Nagar:2011aa}; (ii) their effective 4PN approximation using the
polynomial obtained Taylor-expanding up to formal 4PN the fits of~\cite{Nagar:2011aa} (i.e., using the coefficients listed in Table~I of~\cite{Bernuzzi:2012ku}); (iii) the PN-expanded
$\rho_\lm^H$ obtained from~\cite{Fujita:2014eta}, taken at a PN order that delivers
an excellent agreement with (most of) the numerical data.
In particular, the picture shows the performance of $10$PN for the $\ell=m=2$ mode;
of 15PN   for $\ell = 2, m = 1$; of 6PN for all multipoles with $\ell = 3$, and 
of 12PN for all multipoles with $\ell=4$.
It is interesting to note that for some modes, such as those with $\ell=3$ and $\ell=4$,
the effective 4PN series obtained by expanding the fit is somehow more robust
and accurate, in the strong field, than the high PN expansion.

The impact of these high-order terms on the horizon flux (either the effective ones
or the true PN ones) is explored in Fig.~\ref{fig:5000_newHF}. The top panel 
refers to the $q=5000$ case. The standard curve, with only the $\ell=2$ horizon
flux contributions from Ref.~\cite{Bernuzzi:2012ku}, is contrasted with two different
ways of incorporating more information in the flux. In the first case, we consider 
the PN-expanded numerical fits up to $\ell=4$ (green line). We note that, despite 
the effective nature of the PN coefficients, this choice can already reduce the EOB/GSF disagreement.
In the second case (yellow line) we take advantage of the quality of the PN expansions shown
in Fig.~\ref{fig:rhoH} and use those except for the modes $\ell=m=4$ and $\ell=4, m=2$;
for the latter two modes we prefer to stick to the numerically informed effective PN coefficients due to
the qualitatively different behavior of the PN-expanded functions shown in the right panel of Fig.~\ref{fig:rhoH}\footnote{As a general consideration,
the rather erratic behavior of the PN-expanded $\rho_\lm^H$ indicates that they have
to be additionally resummed. This is usually done for the flux at infinity in \TEOBResumS{}, 
such as in Ref.~\cite{Nagar:2016ayt,Messina:2018ghh}, but it has never been attempted for the horizon
functions in this form (see, however, Ref.~\cite{Taracchini:2013wfa}). Given the importance
of having analytically accurate horizon fluxes, this will be pursued in future work.}.
Figure~\ref{fig:5000_newHF} illustrates that the new, more complete horizon flux lowers
$\Delta \Qo^{\rm EOBGSF}$ by approximately an order of magnitude at $\omega = 0.12$.
By integrating the $\Qo$ difference on the frequency range $\Delta\omega = (0.0224, 0.12)$ for $q = 5000$ 
we find accumulated phase differences of $\sim (0.27, 0.07, -0.01)$ radians for the three 
approximations to the horizon flux. In the bottom panel of the figure we see that the effect is even 
more striking for $q=50000$. The accumulated phase difference up to frequency $\omega=0.12$ 
is halved, from $\sim 5.88$~rad with the standard flux to $\sim 3.07$~rad with the improved flux.

Given that the disagreement still grows with increasing $q$, we can infer that it is still caused by a disagreement in $\Qo^0$.
However, the fact that we no longer see a roughly linear growth with $q$ when moving from $q=5000$ to $q=50000$ 
suggests that the difference $\Delta\Qo^0$ is now sufficiently small that at $q=5000$ it competes with higher-order 
$\Delta \Qo^n$ terms, particularly $\Delta\Qo^1$. The linear growth with $q$ only becomes substantial, and starts to dominate, at high values of $q>5000$.

\section{Conclusions}
\label{sec:conclusions}
We have presented an extensive comparison between a recently proposed EOB model that incorporates linear-in-$\nu$
EOB potentials informed by GSF data~\cite{Nagar:2022fep}  and \PAT{},  a state-of-the-art 2GSF waveform model~\cite{Wardell:2021fyy}. 
We restricted to the quasi-circular case and we have mainly focused on the large-mass-ratio regime, s
o as to investigate the mutual properties of the two models for IMRIs and EMRIs. This study complements 
Paper~I~\cite{Albertini:2022rfe}, which discussed mass ratios up to $q=128$.
Our main findings are as follows:
\begin{itemize}
\item[(i)] We presented EOB/GSF phasing comparisons analogous to those discussed in Paper~I. These rely
           on either time-domain phasing analyses or gauge-invariant phasing analyses based on the $Q_\omega$
           function. We have found that the standard azimuthal radiation reaction implemented in \TEOBResumS{} 
           is insufficient and that it is necessary to incorporate more test-mass terms to achieve an acceptable EOB/GSF 
           waveform agreement.
          In particular, we work at $3^{+19}$PN order in the  residual waveform amplitudes $\rho_\lm$, 
          implementing their high-PN expansions as obtained in Ref.~\cite{Fujita:2012cm}. For simplicity we 
          do not introduce any further resummation of the $\rho_\lm$'s. Also, we sum up modes up to $\ell=8$ and exclude
          the $m=0$ ones\footnote{This is different from \PAT{}, that implements flux modes up to $\ell=30$.}.
           The use of GSF information in both the conservative and nonconservative sectors of the model allows us to
           build  an EOB evolution that is more GSF-faithful for large mass ratios, specifically up to mass ratio $q=500$. 
           This is confirmed both by a time-domain analysis and by a frequency-domain one. Following the same 
           methodology of Paper~I, we have contrasted the coefficients ($\Qo^0, \Qo^1, \Qo^2$) of the $\nu$-expansion of
            $\Qo$ at 2PA, finding an increased EOB/GSF consistency in all three, though mostly in $\Qo^0$ and $\Qo^1$. 
\item[(ii)] To deepen our understanding of the impact of the different contributions to the 2PA $\Qo$, we have expanded 
               the EOB $\Qo$ analytically in $\nu$ for circular orbits, so as to find how $(\FISF, \FIISF, \FIIISF, a_1, a_2)$ enter the three 
               terms ($\Qo^0, \Qo^1, \Qo^2$). This further sheds light on the reason behind the increased EOB/GSF agreement 
               we obtained with the updated EOB model, which is mostly dominated by the (GSF-informed) $(\FISF,a_1)$ functions.
               This also shows that, at least up to $q=500$, the known differences in $\FIISF$ and $\FIIISF$ between \PAT{} and the EOB model
               are not very important, since we find a high degree of consistency {\it also} between the respective $\Qo^2$'s 
               (see in particular the third panel of Fig.~\ref{fig:Qis_w500}). As shown in Paper I, \PAT{} appears to substantially overestimate 
               the true value of $\Qo^2$, suggesting that this consistency in $\Qo^2$ might represent a loss of accuracy in the new EOB model 
               relative to the NR-informed $\Qo^2$ in the standard \TEOBResumS{}; however, this should not be relevant for IMRIs and EMRIs, 
               where $\Qo^2$ makes a very small contribution to the phase.
\item[(iii)] When moving to larger mass ratios, from $q=5000$ to $q=50000$, so as to enter the EMR regime, we have highlighted 
		that a precise modelization of the contribution to the EOB radiation reaction due to the black hole horizon absorption is needed 
		to provide an acceptable EOB/\PAT{} consistency. 
\end{itemize}

Our main general conclusion is that, if properly informed by GSF results (either numerically or analytically), the \TEOBResumS{}
model can generate waveforms that are highly consistent with those generated by \PAT{}. For mass ratios in the hundreds or thousands, once this GSF information is included, we find negligible disagreement between the models over a large frequency interval. Although Fig.~\ref{fig:5000_newHF} shows that there remains a significant disagreement at extreme mass ratios $q\gtrsim 10^4$, our analysis suggests that this can probably be reduced with a further improvement of the infinity and horizon fluxes in the EOB model, specifically in the leading-order fluxes (typically referred to as test-mass fluxes in the EOB literature or as 0PA/1SF fluxes in the GSF literature). Further investigation will be required to determine how best to achieve this improvement, whether by altering the model's resummations or by including more flux multipoles (which are currently truncated at $\ell=8$ in the EOB model, as compared to $\ell=30$ in \PAT{}).

Another important conclusion is that, to a large extent, this refinement of the leading-order fluxes is the main challenge in developing accurate EOB models for EMRIs. Our analyses in Paper I provided additional support to the long-standing belief that EMRI models only require the first two orders in a small-$\nu$ expansion, referred to as 0PA (represented by $\Qo^0$ here) and 1PA (represented by $\Qo^1$); the higher-order coefficients in the expansion are sufficiently well behaved that their suppression by powers of $\nu$ makes them negligible. And as we have shown, an EOB model can achieve good accuracy in 1PA terms by incorporating 1SF information (through the functions $\FISF$ and $a_1$): the impact of the 2SF flux $\FIISF$ appears to be small enough that it may already be sufficiently well captured by the EOB model's baseline representation of the flux, without need for direct information from the GSF calculation of $\FIISF$. Indeed, our comparison shows that the $\Qo^1$ of the updated EOB model already falls within the uncertainty bars of 1PAT1 (see Fig.~1 of Paper I). This suggests that at 1PA order, further investigation is required on the GSF side rather than the EOB side.

In future work, we will explore the EOB model's $\nu$ dependence in more detail. The representations of the fluxes are intrinsically different in the two models, because of a different amount
of $\nu$-dependent information included. \PAT{} calculates the first two orders in the flux,  $\nu^2{\cal F}^{\rm 1SF}_\varphi+\nu^3{\cal F}^{\rm 2SF}_\varphi$, exactly (up to numerical error), while the EOB model approximates these and also includes higher orders in $\nu$, though limited by being based on resummed PN series. In this respect, it might be helpful to build a version of the EOB flux that only includes corrections up to ${\cal F}^{\rm 2SF}_\varphi$, 
though evidently based on resummed PN results up to 3PN. In this way, both \PAT{} and EOB would be 
exactly at the same order in $\nu$, and the only differences should come from the uncertainties in the resummation procedures. 

Once their $\nu$ dependence is clearly delineated, EOB models offer a powerful tool for IMRI and EMRI modelling. An EOB model is clearly {\it very flexible}, as it easily incorporates higher-order-in-$\nu$ terms, either in the radiation 
reaction or in the Hamiltonian. For example, the current EOB model would easily allow us to test, at least approximately, the impact 
of the 2SF correction $a_2$ or 3SF flux $\FIIISF$ on the long inspiral of an IMRI or EMRI, ensuring that the impact is sufficiently small to neglect. The correction $a_2$ is now partly known from PN calculations and 
is among the main challenges of current GSF research. EOB could also be used to inform GSF models, rather than the converse: while 1PAT1 provides a fast, accurate model for quasicircular, nonspinning binaries, GSF models for eccentric, spinning binaries are more limited. Such GSF models include 0PA and some 1PA terms (specifically, 1SF conservative terms) to very high precision but are missing other 1PA terms (specifically, 2SF dissipative terms); an EOB model can provide approximations to those missing terms in regions of the parameter space where 2SF calculations have not yet been performed. Our results in this paper suggest that such approximate 1PA terms may in fact be sufficiently accurate for most purposes, bypassing the need for expensive 2SF calculations, although further work will be needed to assess their accuracy for eccentric, spinning binaries.

It is likely that these mutual synergies between GSF and EOB theory will
be essential in the construction of accurate waveform models for the next generation of detectors.

\acknowledgements
A.A. has been supported by the fellowship Lumina Quaeruntur No.
LQ100032102 of the Czech Academy of Sciences. We are grateful to S.~Albanesi
for discussion and technical help during the development of this work. A.P. gratefully acknowledges the support of a Royal Society University Research Fellowship. 
N.W. acknowledges support from a Royal Society - Science Foundation Ireland University Research Fellowship via grants UF160093 and RGF\textbackslash R1\textbackslash180022.
This work makes use of the Black Hole Perturbation Toolkit \cite{BHPToolkit} and Simulation Tools \cite{SimulationToolsWeb}.

\appendix

\begin{figure*}[t]
\includegraphics[width=0.43\textwidth]{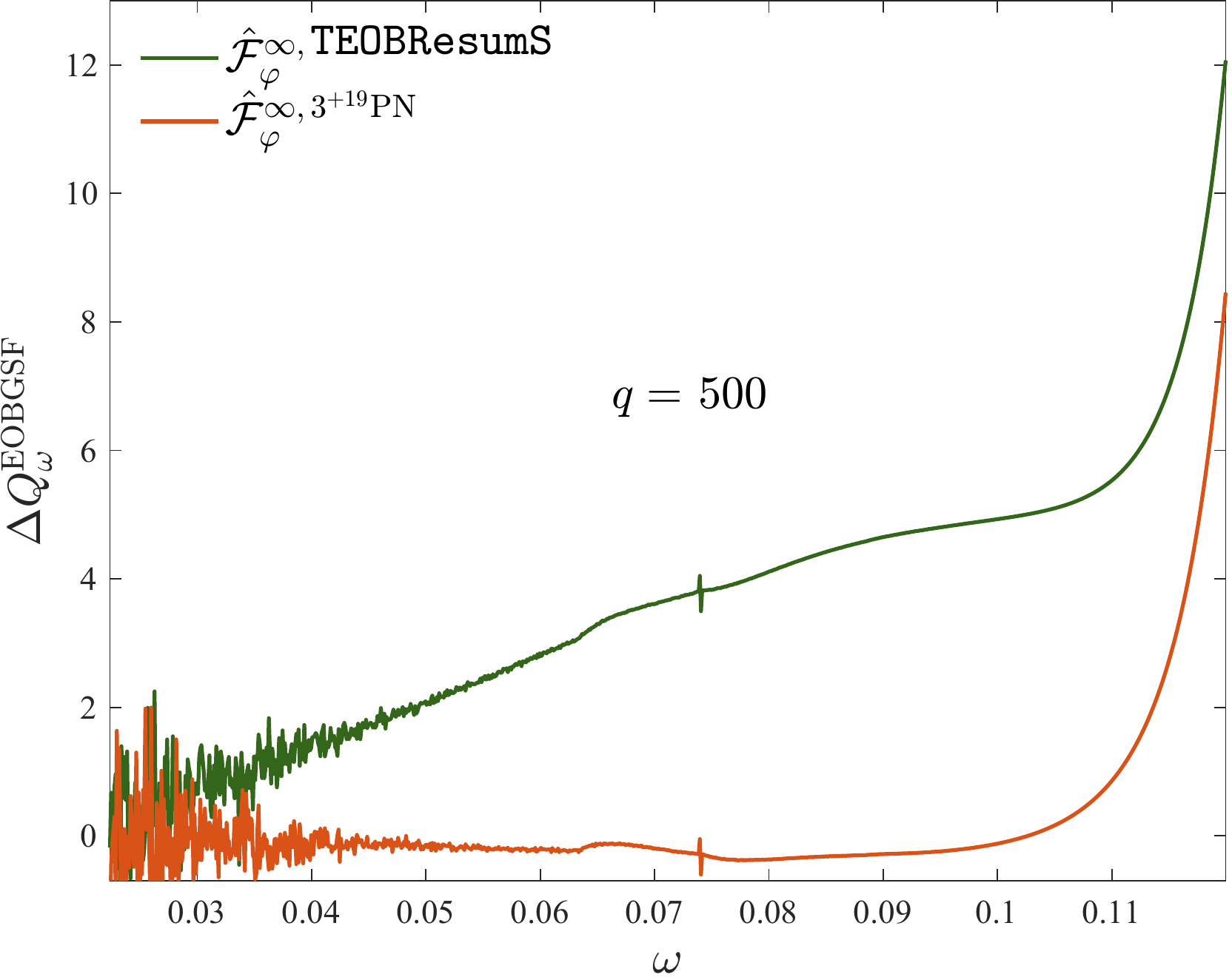} 
\includegraphics[width=0.43\textwidth]{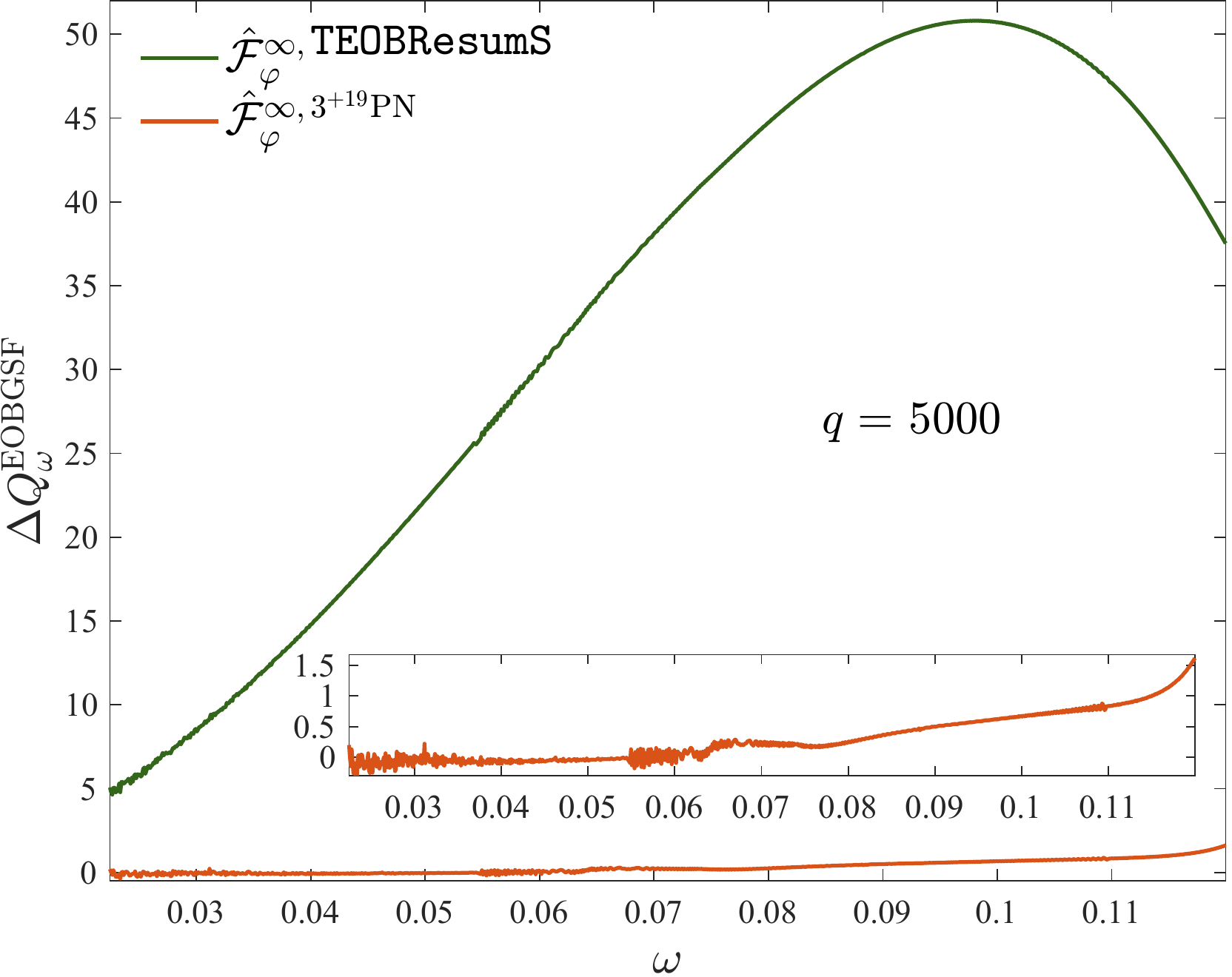} 
\caption{\label{fig:flux_diff}EOB/GSF difference in $\Qo$ for $q = (500, 5000)$, both using the standard \TEOBResumS{} flux for
the evolution and the $3^{+19}$PN flux. Notice how for the second choice the difference starts at zero for both mass ratios.}
\end{figure*}

\section{Inaccuracy of the standard \TEOBResumS{} angular momentum flux}
\label{sec:flux}
In the main text we have mentioned that the standard \TEOBResumS{} flux, as detailed in Ref.~\cite{Nagar:2019wds,Nagar:2020pcj}, 
turns out to be inaccurate as the mass ratio increases and this has a nonnegligible impact on the phasing.
This is testified by Fig.~\ref{fig:flux_diff}, that shows how the EOB/GSF difference is decreased when 
substituting the standard \TEOBResumS{} flux with the $3^{+19}$PN flux described above for mass 
ratios $q = (500, 5000)$. The integrated phase difference up to $\omega = 0.12$ lowers 
from $(4.42, 44.04)$ to $(0.07, 0.27)$ for $q = (500, 5000)$ respectively.

\bibliography{refs20221102.bib, local.bib}

\end{document}